\documentclass{jfm}
\usepackage{natbib}
\usepackage{siunitx}
\begin{document}

\newtheorem{lemma}{Lemma}
\newtheorem{corollary}{Corollary}

\shorttitle{Rotational stabilisation of the {R}ayleigh-{T}aylor instability} %for header on odd pages

\shortauthor{J. Huneault et al.} %for header on even pages

\title{Rotational stabilisation of the {R}ayleigh-{T}aylor instability at the inner surface of an imploding liquid shell}

\author
 {
 Justin Huneault\aff{1}
  David Plant\aff{2}
  \and \\
  Andrew Higgins\aff{1}
  \corresp{\email{andrew.higgins@mcgill.ca}}
  }

\affiliation
{
\aff{1}
Department of Mechanical Engineering, McGill University, 817 Sherbrooke St. West, Montreal, QC H3A 0C3, Canada
\aff{2}
General Fusion Inc., 108-3680 Bonneville Place, Burnaby, BC V3N 4T5, Canada
}

\maketitle

\begin{abstract}
A number of applications utilise the energy focussing potential of imploding shells to dynamically compress matter or magnetic fields, including magnetised target fusion schemes in which a plasma is compressed by the collapse of a liquid metal surface. This paper examines the effect of fluid rotation on the Rayleigh-Taylor (RT) driven growth of perturbations at the inner surface of an imploding cylindrical liquid shell which compresses a gas-filled cavity. The shell was formed by rotating water such that it was in solid body rotation prior to the piston-driven implosion, which was propelled by a modest external gas pressure. The fast rise in pressure in the gas-filled cavity at the point of maximum convergence results in an RT unstable configuration where the cavity surface accelerates in the direction of the density gradient at the gas-liquid interface. The experimental arrangement allowed for visualization of the cavity surface during the implosion using high-speed videography, while offering the possibility to provide geometrically similar implosions over a wide range of initial angular velocities such that the effect of rotation on the interface stability could be quantified. A model developed for the growth of perturbations on the inner surface of a rotating shell indicated that the RT instability may be suppressed by rotating the liquid shell at a sufficient angular velocity so that the net surface acceleration remains opposite to the interface density gradient throughout the implosion. Rotational stabilisation of high-mode-number perturbation growth was examined by collapsing nominally smooth cavities and demonstrating the suppression of small spray-like perturbations that otherwise appear on RT unstable cavity surfaces. Experiments observing the evolution of low-mode-number perturbations, prescribed using a mode-6 obstacle plate, showed that the RT-driven growth was suppressed by rotation, while geometric growth remained present along with significant non-linear distortion of the perturbations near final convergence.
\end{abstract}

\section{Introduction}
\label{sec:1.0}

The influence of the Rayleigh-Taylor instability on implosion phenomena is of relevance to a number of scientific and engineering applications, including $\mathit{Z}$-pinch devices \citep{Golberg1993,Velikovich1996,Ryutov2000}, laser-driven inertial confinement fusion \citep{Hsing1997,Mikaelian2010,Velikovich2015}, and magnetic flux compression \citep{Harris1962,Somon1969,Buyko1997}. Of particular interest for the present work is a magnetised target fusion (MTF) \citep{Kirkpatrick1995,Buyko1997} concept in which a plasma target is compressed by an imploding liquid metal surface to reach fusion conditions \citep{Laberge2008,Turchi2008,Suponitsky2014}. In this concept, which was initially proposed during the LINUS program \citep{Turchi1980,Robson1982}, a cylindrical or spherical rotating liquid shell is collapsed by mechanical pistons in a quasi-reversible cycle in order to compress the plasma to fusion conditions. The resulting energy release is deposited into the shell in the form of heat and radial kinetic energy. The mechanical pistons are re-compressed by the rebounding liquid shell, creating a reversible scheme in which excess energy can be extracted from the shell via a heat exchanger \citep{Robson1982}. The pressure limitations on mechanical systems restricts the collapse velocity to the range of \SIrange{e2}{e3}{\metre\per\second}, which results in implosion time scales on the order of hundreds of microseconds to milliseconds \citep{Robson1982,Laberge2008}. To reach the conditions necessary for breakeven fusion burn, perturbations on the inner wall of the liquid metal shell must not be allowed to grow to sufficient size such that they cause mixing of the cool wall material with the plasma or limit convergence by encroaching onto the central axis of the implosion. The former is a concern for high-mode-number (small wavelength) perturbations, where small jets or drops may be vaporised by the magnetic field, while the latter is primarily affected by the growth of large, low-mode-number perturbations \citep{Somon1969}.

It is well known that the Rayleigh-Taylor (RT) instability \citep{Rayleigh1883,Taylor1950}, which occurs when the interface between fluids of different densities is accelerated in the direction of the heavy fluid, has significant consequences for fusion concepts which rely on implosion, such as inertial confinement fusion (ICF) \citep{Sweeney1981,Betti1998}. In ICF, a cryogenic deuterium target is compressed by a thin shell imploded via laser-driven ablation on a nanosecond timescale. The inward push from the laser-ablated material during the acceleration stage causes RT-driven perturbation growth at the outer surface of the shell. These perturbations feed-through to the inner surface and undergo further perturbation growth near the point of maximum compression when the shell inner surface is decelerated by the light deuterium target \citep{Weir1998,Wang2015}. The use of mechanical pistons to drive the implosion in the MTF scheme discussed above ensures a hydrodynamically stable outer surface, but it remains likely that perturbations will form on the inner surface of the liquid shell due to flow obstacles in the mechanical system or surface waves that result from the initial flow needed to form the cylindrical or spherical cavity. At the point of maximum compression (i.e., \textit{turnaround}), the shell kinetic energy is rapidly transferred to the plasma target, resulting in a large increase in pressure and subsequent deceleration of the shell inner surface. At this stage of the implosion, the radial acceleration of the cavity inner surface is aligned with the density gradient at the plasma-liquid interface, which causes the growth of perturbations due to the RT instability. An analogous phenomenon has been observed in experiments where magnetic fields are compressed by an imploding metal shell \citep{Fowler1960,Alikhanov1968}, where it is believed that the magnetic field amplification is limited by the growth of perturbations near turnaround, due to the deceleration of the liquefied shell by the magnetic field target \citep{Cnare1966,Somon1969}. 

The applications listed above have motivated the development of models for the growth of perturbations at cylindrical \citep{Bell1951,Mikaelian2005} and spherical \citep{Bell1951,Plesset1954,Mikaelian1990} interfaces subject to the RT instability. The time ($t$) evolution of an azimuthally perturbed cylindrical interface between a light inner fluid of density $\rho_1$ and a heavy outer fluid of density $\rho_2$ can be described by the second-order linear differential equation below \citep{Mikaelian2005}

\begin{equation}
\label{eq:1:1}
0=\ddot{\eta}+2\frac{\dot{R}_1}{R_1}\dot{\eta}-\frac{\ddot{R}_1}{R_1}(M\mathcal{A} -1)\eta,
\end{equation}

\noindent where $\mathcal{A}$ is the Atwood number, a non-dimensional parameter defined below which describes the density ratio between the fluids.

\begin{equation}
\label{eq:1:2}
\mathcal{A}=\frac{\rho_2-\rho_1}{\rho_2+\rho_1},
\end{equation}

\noindent $R_1$, $\dot{R}_1$, and $\ddot{R}_1$ are the interface radius and its time derivatives, $\eta$, $\dot{\eta}$, and $\ddot{\eta}$ are the amplitude of the sinusoidal perturbations and their time derivatives, and $M$ is the mode number of the sinusoidal perturbation. The model, which applies to small amplitude perturbations, assumes the fluids are incompressible and that the flow is irrotational. Equation~\ref{eq:1:1} shows that in a cylindrical geometry, perturbation growth is coupled to the evolution of the radius, velocity, and acceleration of the interface. Of particular interest is the contribution of the interface acceleration ($\ddot{R}_1$ term), which induces exponential perturbation growth for positive (outward facing) accelerations due to the RT instability, or stable oscillations (i.e., perturbation phase reversals) for negative interface accelerations. Convergence ($R_1$ decreasing) effects, known as Bell-Plesset effects due to their seminal studies of the evolution of a perturbed imploding cylindrical or spherical interface \citep{Bell1951,Plesset1954}, cause perturbations to grow as implosion progresses even for an RT stable interface. From equation~\ref{eq:1:1}, it can be seen that for an RT unstable interface, the perturbation amplitude is expected to grow exponentially with a growth rate described by the equation below \citep{Epstein2004}.

\begin{equation}
\label{eq:1:3}
\gamma=\bigg(\frac{M}{R_1}\ddot{R}_1 \mathcal{A}\bigg)^{\frac{1}{2}},
\end{equation}

\noindent This growth rate is valid in the linear regime, where the amplitude of the perturbation is below approximately 10\% of the wavelength ($\eta<0.2\upi R_1/M$), at which point nonlinear saturation causes a transition to linear growth \citep{Haan1989,Hsing1997}. As expected, the growth rate of the perturbations increases as the interface acceleration becomes greater. The dependence of growth rate on wavenumber ($k=M/R_1$) results in small wavelength perturbations growing faster than larger ones. This phenomenon is ultimately limited by surface tension, which stabilises small wavelength perturbations \citep{Chandrasekhar1961}. 

Interest in stabilising fluid interfaces has motivated significant research on the RT instability in rotating systems. The effect of rotation on a planar fluid interface normal to the rotation axis has been studied theoretically \citep{Chandrasekhar1961,Scase2017} and experimentally \citep{Baldwin2015}. Although rotation does not stabilise the interface \citep{Chandrasekhar1961}, the tendency of a rotating fluid to form Taylor columns inhibits relative lateral movement between the fluids at the interface, which suppresses the growth of large-scale perturbations \citep{Baldwin2015}. The present paper focuses on the effect of rotation at a cylindrical fluid interface parallel to the axis of rotation, where the centripetal acceleration at the interface has an important effect on stability. It has been proposed that the RT-driven growth of perturbations near turnaround in imploding shell drivers for magnetic field compression or MTF applications may be mitigated by rotating the shell prior to implosion \citep{Barcilon1974}. In this concept, the inward pointing centripetal acceleration at the shell inner surface offsets the large outward radial acceleration near turnaround, allowing the net surface acceleration to remain in the opposite direction to the density gradient throughout the implosion. If the shell is assumed to behave as an inviscid liquid, the angular velocity of the shell inner surface near maximum compression is determined by the initial rate of solid body rotation and the significant increase in angular velocity that results from the conservation of angular momentum during the implosion. As will be seen, the initial angular velocity required to stabilise the inner surface of the shell from RT instability growth is determined by a number of factors including the size, thickness, and density of the shell, the properties of the target gas or magnetic field, and the outer pressure acting on the shell. Models developed for the perturbed motion of an imploding incompressible and inviscid shell with an outer (driving) and inner (target) magnetic field have shown that rotation can reduce the growth of perturbations near turnaround \citep{Barcilon1974}, albeit at the cost of reduced payload compression \citep{Book1974}. These models were extended to study cases of high velocity or high convergence ratio implosions where compressibility must be taken into account \citep{Book1979a,Book1979b}. Early experiments related to the LINUS program demonstrated an improvement in the stability of the inner surface of magnetically driven shells with increased angular velocity \citep{Turchi1976}. The rotational stabilisation technique was extended to a mechanical-piston-driven implosion device capable of collapsing water liners over a radial compression of 30:1 while maintaining an optically smooth inner surface \citep{Turchi1980}. Recently, \citet{Avital2019} developed an analytical model for the motion of perturbations on the inner surface of a cylindrically imploding liner using an approach similar to that of \citet{Barcilon1974}, which also showed that rotation can supress RT-driven perturbation growth near turnaround, a result that was corroborated by CFD simulations. The effect of rotation on the growth of low-mode-number perturbations considered in the modelling work of \citet{Barcilon1974} and \citet{Avital2019} has not been explored experimentally.

This paper will examine the effect of fluid rotation on the growth of perturbations at the inner surface of an imploding cylindrical liquid-water shell initially in solid body rotation. In section~\ref{sec:4.0}, a model for perturbation growth on the inner surface of a rotating shell will be developed. The experimental arrangement for collapsing rotating shells will be presented in section~\ref{sec:2.0}, followed by the development and experimental validation of a model for the unperturbed one-dimensional motion of the cavity surface in section~\ref{sec:3.0}. Finally, sections~\ref{sec:5.0} and~\ref{sec:6.0} will present and discuss the experimental results and compare them to the perturbation growth model.

\section{Model for small amplitude perturbation growth on a cylindrically imploding cavity within a rotating fluid}
\label{sec:4.0}

\begin{figure}
	\centering
	\includegraphics[width=0.3\columnwidth]{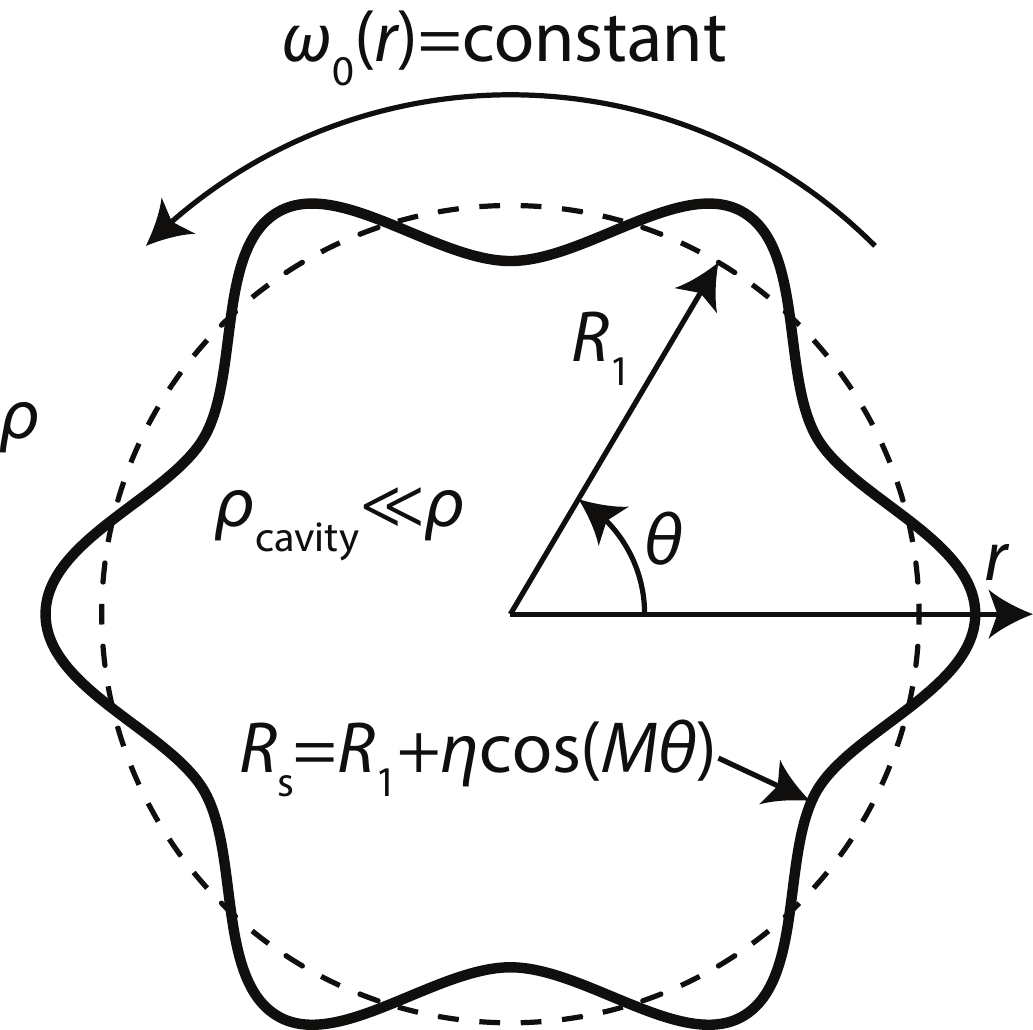}
	\caption{Schematic of the problem definition for the development of the model for the perturbation growth on the inner surface of a rotating cylindrically imploding fluid.}
	\label{fig:1}
\end{figure}

This section will present the development of an ordinary differential equation for the time ($t$) dependent variation in the amplitude of azimuthal perturbations ($\eta$) on a cylindrical interface, with time varying radius ($R_1(t)$ not constant), between an inner compressible fluid of negligible density and an outer inviscid fluid initially in solid body rotation. The problem is illustrated schematically in figure~\ref{fig:1}. The derivation will follow the methodology of previous authors \citep{Bell1951,Plesset1954,Mikaelian1990,Mikaelian2005} who have studied the RT instability in non-rotating converging geometries, with the addition of a gravitational potential ($\xi$) term to the equation of motion that accounts for the effect of rotation on the pressure distribution within the outer fluid. Treating the effect of rotation using a gravitational potential, rather than solving for the rotational flow at the perturbed interface, significantly simplifies the model. This approach allows for the motion of the azimuthal perturbations on the moving interface to be defined by a simple linear second-order differential equation that resembles the original non-rotating derivation of Bell \citep{Bell1951}. The treatment therefore neglects the effect of the Coriolis acceleration on the perturbed flow, a factor that will be discussed in detail in section~\ref {sec:6.0}. In determining the rotation gravitational potential field, it will be assumed that the angular momentum of the fluid is conserved during the implosion, which causes the angular velocity of the fluid at the interface to increase as the cavity collapses. The motion of the perturbations will be treated assuming an inviscid potential flow, while the treatment of the perturbed values relies on linearisation of the equations, as such the analysis will only apply to small-amplitude perturbations. The inviscid outer fluid is assumed to be incompressible, while the inner fluid, which is not considered in the analysis due to its negligible density, is thus assumed to be fully compressible. The analysis is performed in the high Atwood number limit ($\mathcal{A}\approx 1$ for $\rho_{\mathrm{cavity}}\ll\rho$), which is applicable to the experiments considered in this study as well as MTF applications, where a magnetized plasma is compressed by an imploding metal wall. The effect of surface tension ($\sigma$) at the interface has been neglected, which limits the applicability to relatively low wavenumber perturbations: in the high Atwood number limit, the effect of surface tension on the RT instability is negligible when $k \ll (g\sigma /\rho)^{0.5}$ \citep{Chandrasekhar1961}, where $g$ is the interface acceleration. The assumption is valid for the mode-6 perturbations that will be considered in this work. The outer fluid is assumed to extend to infinity, and as a result, the feedthrough of perturbations from the outer surface are not considered. This assumption is justified considering the thickness of the shell and the stable implosion drive mechanism used in this study. Sinusoidal perturbations along the azimuthal direction ($\theta$) are introduced to the fluid interface, such that the position and velocity of the perturbed interface ($R_\mathrm{s}$) can be defined by

\begin{equation}
\label{eq:4:1}
R_ {\mathrm{s}}=R_1+\eta \mathrm{cos}(M\theta),
\end{equation}

\begin{equation}
\label{eq:4:2}
\dot{R}_ {\mathrm{s}}=\dot{R}_1+\dot{\eta}\mathrm{cos}(M\theta),
\end{equation}

\noindent where $R_ 1$ is the nominal radius of the interface, $\eta$ is the perturbation amplitude, and $M$ is the mode number of the perturbation. A velocity potential ($\varphi$) for the radial flow within the outer fluid can be defined such that

\begin{equation}
\label{eq:4:3}
\dot{r}=-\Big(\frac{\partial \varphi}{\partial r}\Big),
\end{equation}

\noindent where $r$ denotes the radial position within the outer fluid and $\dot{r}$ the radial velocity. For an incompressible irrotational cylindrically imploding flow $\nabla^2 \varphi =0$ and from continuity $r\dot{r}=R_1\dot{R}_1$, so that the velocity potential for the outer fluid can be defined as \citep{Bell1951,Mikaelian2005}

\begin{equation}
\label{eq:4:4}
\varphi=-R_1\dot{R}_1\mathrm{ln}(r) + \frac{b}{r^M}\mathrm{cos}(M\theta),
\end{equation}

\noindent where $b$ is a time-dependent function that is independent of the radial position. The velocity potential has been defined in this manner such that the disturbance decays moving away from the interface. The function $b$ can be determined by applying the definition of the velocity potential in equation~\ref{eq:4:3} at the fluid interface, where the radial velocity must be equal to the velocity of the boundary:

\begin{equation}
\label{eq:4:5}
-\Big(\frac{\partial\varphi}{\partial r}\Big)_{r=R_1+\eta\mathrm{cos}(M\theta)}=\dot{R}_1+\dot{\eta}\mathrm{cos}(M\theta).
\end{equation}

\noindent Performing the spatial partial derivative:

\begin{equation}
\label{eq:4:6}
\Big[\frac{R_1\dot{R}_1}{r}+\frac{Mb}{r^{M+1}}\mathrm{cos}(M\theta)\Big]\Big|_{r=R_1+\eta\mathrm{cos}(M\theta)}=\dot{R}_1+\dot{\eta}\mathrm{cos}(M\theta).
\end{equation}

\noindent In order to solve equation~\ref{eq:4:6}, the following linear approximations will be made:

\begin{equation}
\label{eq:4:7}
\frac{1}{r}\Big|_{R_1+\eta\mathrm{cos}(M\theta )}\approx \frac{1}{R_1}\Big(1-\frac{\eta\mathrm{cos}(M\theta)}{R_1}\Big),
\end{equation}

\begin{equation}
\label{eq:4:8}
\frac{1}{r^{M+1}}\Big|_{R_1+\eta\mathrm{cos}(M\theta )}\approx \frac{1}{R_1^{M+1}}.
\end{equation}

\noindent As was noted by \citet{Mikaelian1990}, the linearisation in equation~\ref{eq:4:7} requires that $|\eta|~\ll~R_1$, while equation~\ref{eq:4:8} requires the more stringent condition that $M|\eta|~\ll~R_1$. Equation~\ref{eq:4:6} can now be solved for $b$:

\begin{equation}
\label{eq:4:9}
b=\frac{R_1^{M+1}}{M}\big(\dot{\eta}+\eta\frac{\dot{R}_1}{R_1}\big).
\end{equation}

\noindent Substituting the expression for $b$ into equation~\ref{eq:4:4}, the radial velocity potential within the outer fluid can then be written as follows:

\begin{equation}
\label{eq:4:10}
\varphi=-R_1\dot{R}_1\mathrm{ln}(r) + \frac{R_1^{M+1}}{M\, r^M}\big(\dot{\eta}+\eta\frac{\dot{R}_1}{R_1}\big)\mathrm{cos}(M\theta).
\end{equation}

\noindent Rotation will be treated in the equation of motion using a gravitational potential term which takes into account the effect of rotation on the pressure distribution within the liquid shell. The gravitational potential ($\xi$) will be calculated using the unperturbed shell radius ($R_1$) as a datum. The expression for the gravitational potential will be derived for the specific case of a fluid initially in solid body rotation, and will take into account the gradient in the angular velocity within the shell to first order. Furthermore, it will be assumed that the fluid is inviscid, resulting in conservation of angular momentum as the interface is displaced. The gravitational potential as a function of radial position ($r$) in the fluid can be found using the following integral:

\begin{equation}
\label{eq:4:11}
\xi(r)=\int_{R_1}^{r}r\omega^2dr,
\end{equation}

\noindent where $\omega$ is the local angular velocity of the fluid and must be determined from the initial rate of solid body rotation. From angular momentum conservation, $\omega$ can be obtained from the initial angular velocity of the fluid ($\omega_0$) and the initial radial position of the fluid ($r_0$):

\begin{equation}
\label{eq:4:12}
\omega=\omega_0\Big(\frac{r_0}{r}\Big)^2,
\end{equation}

\noindent where the initial radial position of the fluid can be determined from the initial ($R_{1_0}$) and current ($R_\mathrm{s}$) positions of the fluid interface using the conservation of mass:

\begin{equation}
\label{eq:4:13}
r_0^2=r^2-R_\mathrm{s}^2+R_{1_0}^2.
\end{equation}

\noindent Substituting equations~\ref{eq:4:12} and~\ref{eq:4:13} into equation~\ref{eq:4:11} and performing the integration, the following expression for the gravitational potential term can be obtained:

\begin{equation}
\label{eq:4:14}
\xi(r)=\omega_0^2\Big[\frac{r^2-R_1^2}{2}+2\mathrm{ln}\Big(\frac{r}{R_1}\Big)(R_{1_0}^2-R_\mathrm{s}^2)-\frac{1}{2}\Big(\frac{1}{r^2}-\frac{1}{R_1^2}\Big)(R_\mathrm{s}^4+R_{1_0}^4-2R_\mathrm{s}^2R_{1_0}^2)\Big].
\end{equation} 

An equation of motion for the outer fluid can now be written by considering the Bernoulli equation for unsteady potential flow:

\begin{equation}
\label{eq:4:15}
\frac{\partial\varphi}{\partial t} - \frac{1}{2}(\nabla\varphi)^2 + \frac{P}{\rho} + \xi= F(t),
\end{equation}

\noindent where $F(t)$ is an integration constant or Bernoulli constant. Following the approach of previous workers \citep{Bell1951,Plesset1954,Mikaelian1990}, equation~\ref{eq:4:15} will be evaluated at the fluid interface ($r=R_1+\eta\mathrm{cos}(M\theta )$). The linear terms proportional to $\eta$ (i.e., dependent on $\theta$) in equation~\ref{eq:4:15} will then be collected, resulting in an equation for the time evolution of the perturbation amplitude. First, as noted by \citet{Bell1951}, the pressure $P$ is assumed to be uniform (independent of $\theta$) at the interface and, therefore, does not need to be determined. Similarly, the integration constant $F(t)$, which represents the pressure at infinity \citep{Plesset1954}, is also independent of $\theta$ and does not need to be solved for. The partial time derivative of the velocity potential at the interface can readily be obtained:

\begin{eqnarray}
\label{eq:4:16}
\Big(\frac{\partial\varphi}{\partial t}\Big)=-\mathrm{ln}(r)\frac{\mathrm{d}}{\mathrm{d}t}(R_1\dot{R}_1)+\frac{(M+1)R_1^M\dot{R}_1\dot{\eta}}{Mr^M}\mathrm{cos}(M\theta)+\frac{R_1^{M+1}\ddot{\eta}}{Mr^M}\mathrm{cos}(M\theta) \nonumber\\
+\frac{R_1^{M-1}\dot{R}_1^2\eta}{r^M}\mathrm{cos}(M\theta)+\frac{R_1^M\dot{R}_1\dot{\eta}}{Mr^M}\mathrm{cos}(M\theta)+\frac{R_1^M\ddot{R}_1\eta}{Mr^M}\mathrm{cos}(M\theta),
\end{eqnarray}

\begin{eqnarray}
\label{eq:4:17}
\Big(\frac{\partial\varphi}{\partial t}\Big)\Big|_{r=R_1+\eta\mathrm{cos}(M\theta )}=-\mathrm{ln}(R_1)\frac{\mathrm{d}}{\mathrm{d}t}(R_1\dot{R}_1)+\frac{M+2}{M}\dot{R}_1\dot{\eta}\mathrm{cos}(M\theta) \nonumber\\
+\frac{1}{M}R_1\ddot{\eta}\mathrm{cos}(M\theta)+\frac{1-M}{M}\ddot{R}_1\eta\mathrm{cos}(M\theta),
\end{eqnarray}

\noindent where we have used $\mathrm{ln}(r)|_{R_1+\eta\mathrm{cos}(M\theta )}\approx\mathrm{ln}(R_1)+\eta\mathrm{cos}(M\theta )/R_1$ to linearized the natural logarithm term. The square of the gradient of the velocity potential is equal to the square of the derivative of the velocity potential in the radial direction to first order, which from equation~\ref{eq:4:3}, is equal to the square of the radial velocity at the interface:

\begin{equation}
\label{eq:4:18}
(\nabla\varphi)^2|_{r=R_1+\eta\mathrm{cos}(M\theta )}\approx\Big(\frac{\partial\varphi}{\partial r}\Big)^2|_{r=R_1+\eta\mathrm{cos}(M\theta )}=\dot{R}_1^2+2\dot{R}_1\dot{\eta}\mathrm{cos}(M\theta),
\end{equation}

\noindent keeping only terms linear in $\eta$. As \citet{Plesset1954} noted, the contribution of the velocity component in the $\theta$ direction to $(\nabla\varphi)^2$ is of second order and can be neglected. The expression for the gravitational potential can now be determined by evaluating equation~\ref{eq:4:14} at the interface:

\begin{equation}
\label{eq:4:19}
\xi\big|_{r=R_1+\eta\mathrm{cos}(M\theta )}=\omega_0^2\frac{R_{1_0}^4}{R_1^3}\eta\mathrm{cos}(M\theta),
\end{equation} 

\noindent where only first order terms in $\eta$ have been kept by using previously mentioned linearized expressions as well as $1/{r^2}|_{R_1+\eta\mathrm{cos}(M\theta)}\approx1/R_1^2(1-2\eta\mathrm{cos}(M\theta)/R_1)$. The $\omega_0^2R_{1_0}^4/R_1^3$ factor is easily recognized as the centripetal acceleration of the fluid at the nominal interface radius. The terms from equations~\ref{eq:4:17}-\ref{eq:4:19} can now be substituted into equation~\ref{eq:4:15}. Collecting the terms proportional to $\eta$, the following relation can be obtained:

\begin{equation}
\label{eq:4:20}
0=\ddot{\eta}+2\frac{\dot{R}_1}{R_1}\dot{\eta}-\frac{1}{R_1}(-M a_\mathrm{net}-\ddot{R}_1)\eta,
\end{equation} 

\noindent where $a_\mathrm{net}$ is the net surface acceleration, defined as $a_\mathrm{net}=-(\ddot{R}_1-\omega_0^2 R_{1_0}^4/R_1^3)$ such that it is positive for an inward-facing acceleration. Equation~\ref{eq:4:20} is a second-order differential equation that describes the time evolution of small perturbations on the cylindrical interface between an outer fluid initially in solid body rotation and an inner fluid of negligible density. As can be seen, the final term of equation~\ref{eq:4:20} (proportional to $\eta$), which is responsible for the RT-driven growth of perturbations (for a positive $\ddot{R}_1$), is stabilised by the centripetal acceleration of the fluid. The growth rate of RT-instability-driven perturbation growth in equation~\ref{eq:4:20} can be expressed as

\begin{equation}
\label{eq:4:21}
\gamma=\bigg(\frac{M}{R_1} (-a_\mathrm{net})\bigg)^{\frac{1}{2}},
\end{equation}

\noindent for $a_\mathrm{net} < 0$. When comparing equation~\ref{eq:4:21} to equation~\ref{eq:1:3} for an Atwood number of 1, which treats the same problem considered here but without the gravitational potential induced by rotation, it can be seen that for an otherwise RT unstable interface, where $\ddot{R}_1$ is positive, the centripetal acceleration ($\omega_0^2R_{1_0}^4/R_1^3$) reduces the growth rate of the RT instability and can entirely stabilise the interface if $a_\mathrm{net}$ remains positive. The model derivation presented above has demonstrated that the centripetal acceleration should be expected to reduce the growth rate of RT-driven perturbations on the surface of a rotating imploding shell, and entirely stabilises the interface from the RT instability provided that the net acceleration remains inward facing throughout the implosion. 

It is important to reiterate that the model derived above does not consider the effect of the Coriolis acceleration on the perturbed flow. Rotation affects the perturbed interface through two mechanisms: 1. The azimuthal component of the Coriolis acceleration, which is responsible for the conservation of angular momentum, causes an angular velocity gradient across the perturbations ($\omega \sim 1/r^2$) that distorts the perturbations. 2. Perturbation growth and bulk cavity motion ($\dot{\eta}$ and $\eta\dot{R}_1/R_1$) induce a flow in the azimuthal direction, which generates a Coriolis acceleration in the radial direction that has been shown to stabilise a perturbed interface \citep{Tao2013}. These effects will be discussed further in section~\ref{sec:6.0}.

\section{Apparatus description}
\label{sec:2.0}

\begin{figure}
  \centerline{\includegraphics[width=1\columnwidth]{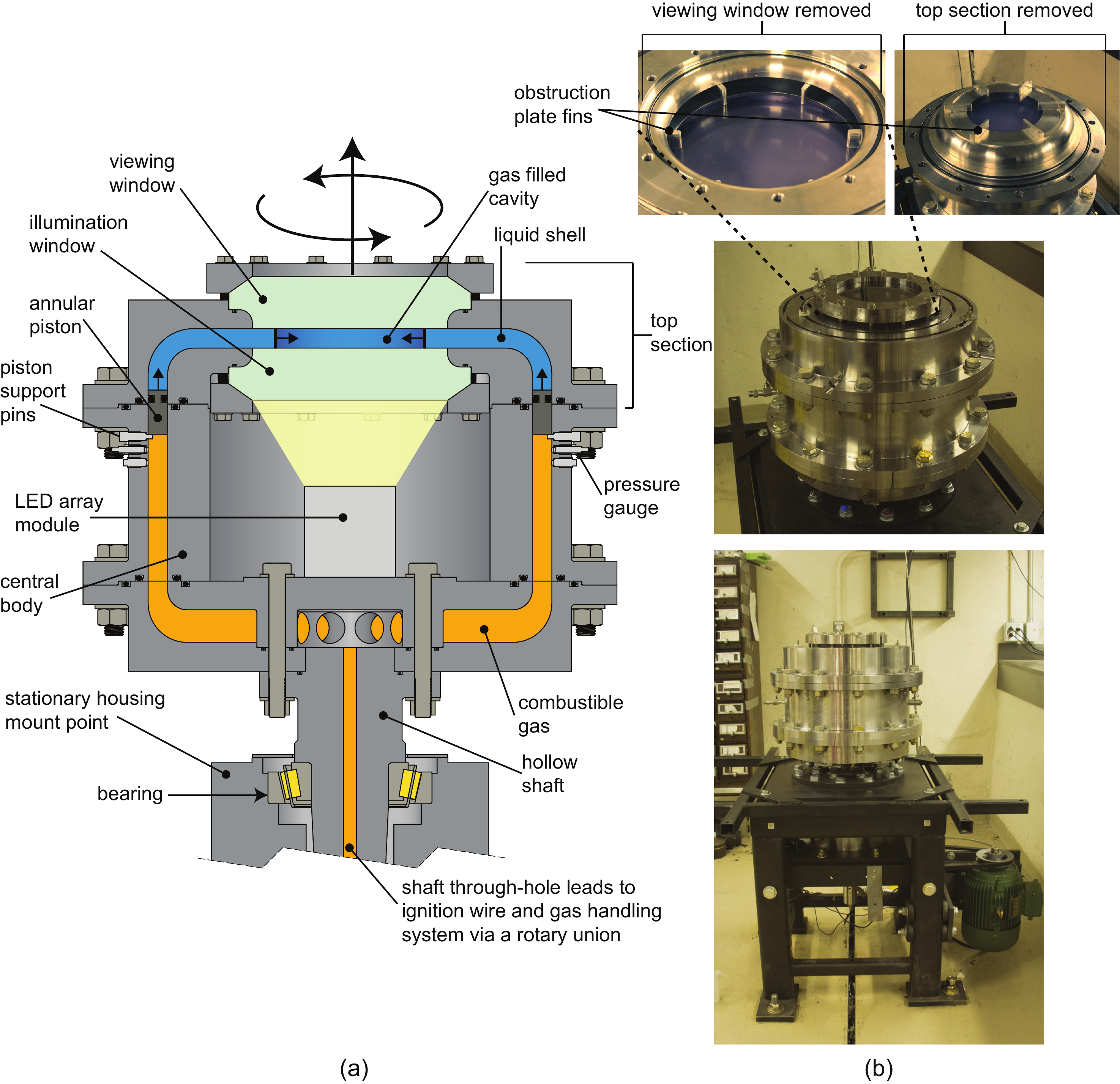}}
  \caption{Cylindrical shell implosion apparatus (a) a labelled section view schematic, (b) pictures of the assembled apparatus, including (top) a picture with the viewing window removed to expose the six-fin obstruction plate.}
  \label{fig:2}
\end{figure}

\begin{figure}
  \centerline{\includegraphics[width=0.5\columnwidth]{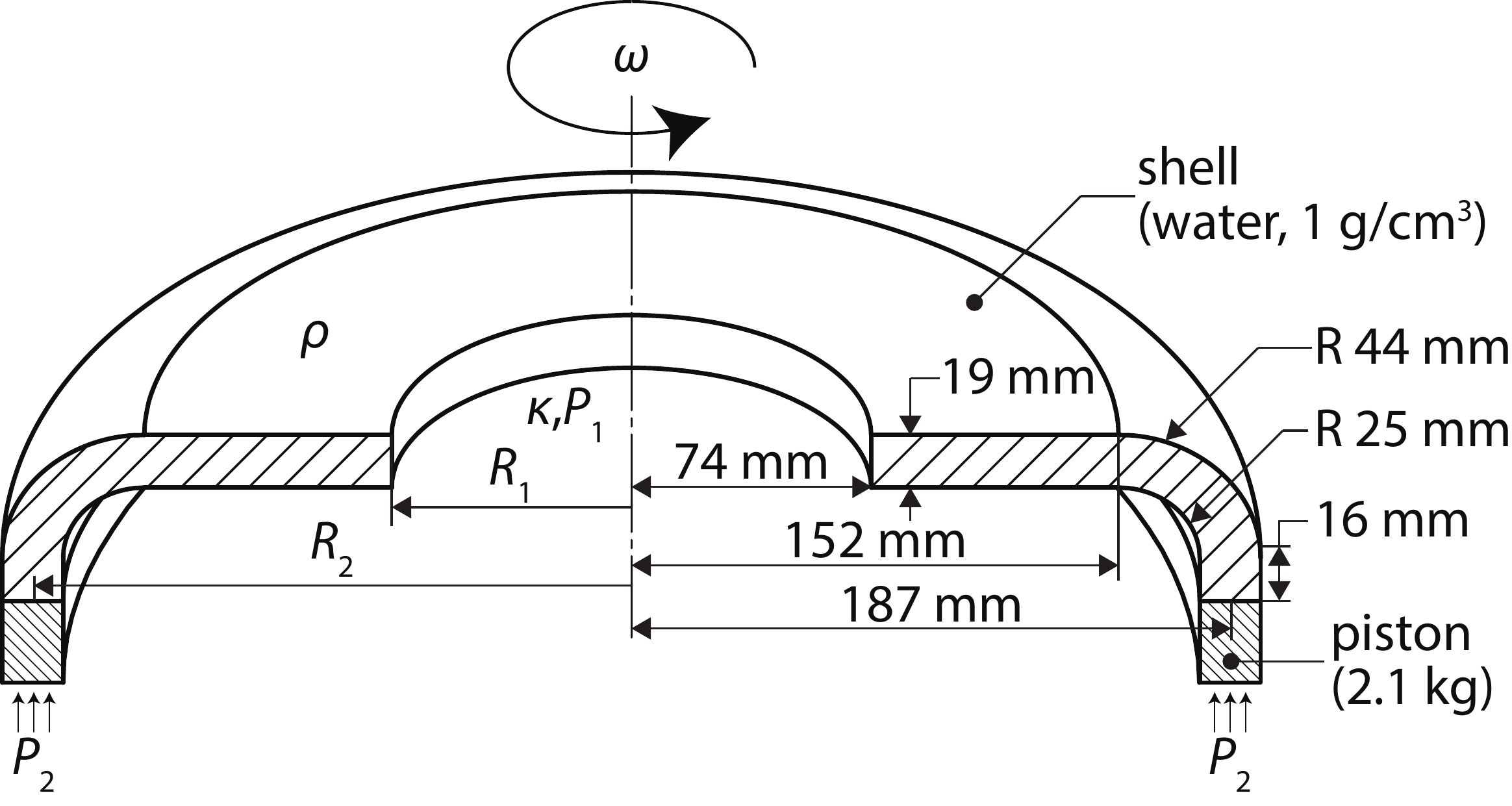}}
  \caption{Schematic of the liquid shell and piston contained within the apparatus with key dimensions.}
  \label{fig:3}
\end{figure}

The experiments presented in the following sections were performed using a rotating cylindrical pressure vessel, which is shown both with a section-view schematic and in pictures in figure~\ref{fig:2}. The shaft-mounted vessel is supported by a bearing housing that is connected to a steel frame, allowing the entire apparatus to rotate. A central body placed within the cylindrical vessel separates the apparatus into a top and bottom cylindrical section connected via curved transition sections to a long annular channel. The implosion of the liquid shell occurs in the top semi-cylindrical section, which is partially filled with water prior to rotation. The water is vacuum degassed in-situ to remove trapped gas bubbles and the section is pressurised, typically with ethane gas, to the desired initial cavity gas pressure. As the apparatus is rotated via a 3~hp electric motor, the water is forced to the outside, thus forming a quasi-cylindrical shell with a gas filled cavity in its center. A dimensioned schematic of the liquid shell is shown in figure~\ref{fig:3}. The liquid shell is held on its outer surface by an o-ring sealed annular piston which rests on a set of pins until the implosion is initiated. The annular channel and cylindrical section below the piston are pre-filled with combustible gas through a hole in the shaft that connects to the gas handling system via a rotary union. The implosion is initiated at the bottom of the shaft by igniting the combustible gas mixture using a heated tungsten wire held within the rotary union. A relatively insensitive mixture of stoichiometric methane-oxygen diluted with nitrogen gas (CH$_4$ + 2 O$_2$ + 3.7 N$_2$) is used to avoid detonation of the gas within the confined pressure vessel, thus allowing for a gradual ramp-up in pressure which can be tuned by modifying the initial combustible gas pressure. The implosion is driven by a moderate pressure (≈1 MPa) resulting in implosion timescales on the order of milliseconds. Viewing windows at the top and bottom of the cylindrical channel allow for visualisation of the cavity collapse. The central body is made hollow such that it can contain a high-power LED lighting module providing backlighting to the experiment. As can be seen in figure~\ref{fig:2}, obstruction plates can be placed within the top cylindrical section to impose perturbations on the surface of the cavity.

\begin{figure}
  \centerline{\includegraphics[width=0.75\columnwidth]{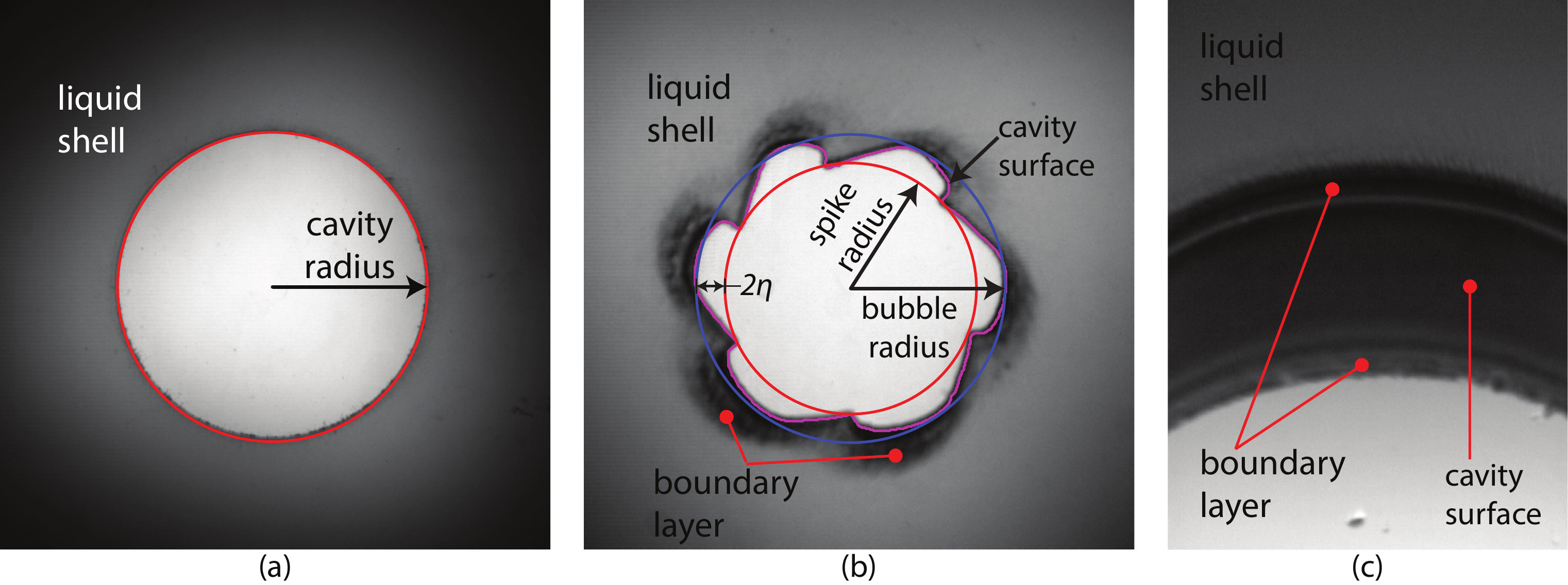}}
  \caption{Sample images taken during experiments. (a) and (b): examples of the image processing algorithm for (a) unperturbed experiments with a uniform radius and (b) experiments with seeded mode-6 perturbations. (c): sample off-axis image for an unperturbed experiment, where the banding caused by the boundary layer at the top and bottom of the test section has been labelled.}
  \label{fig:4}
\end{figure}

The cavity implosions were visualised using two synchronised high-speed cameras. A normal (i.e., on-axis) camera angle was used to track the radial position of the cavity and measure the evolution in the cavity radius over time as well as the growth of low-mode-number seeded perturbations. An off-axis view was used to view the cavity surface and observe the appearance of high-mode number spray-like perturbations on the cavity surface. A Photron FASTCAM SA5 was used for capturing the normal view perspective, while a Photron FASTCAM SA1.1 was used for the off-axis imaging. An edge-tracking algorithm that relies on the change in light intensity between the cavity and the liquid shell was used to extract the radial position of the cavity and the amplitude of the low-mode-number seeded perturbations. The water which formed the shell was died with food colouring to increase the light contrast at the edge. Sample images of the edge-tracking algorithm for unperturbed and perturbed implosions are shown in figure~\ref{fig:4}. As can be seen, the amplitude ($\eta$) of the seeded perturbations is defined as half of the difference between the bubble and spike radius. A sample off-axis image is shown in figure~\ref{fig:4}(c), where a nominally smooth cavity surface can be observed between the top and bottom edges. The effect of the boundary layer that forms on the top and bottom of the relatively thin test section (19~mm) can be observed in both the on-axis and off-axis images. Figure~\ref{fig:4}(c) shows that the boundary layer causes banding near the top and bottom edge of the cavity, but the central portion of the interface is not affected by the velocity shear at the windows. As can be seen in figure~\ref{fig:4}(b), the boundary layer forms dark areas near the cavity surface of the on-axis images, particularly near turnaround where the velocity gradients are large and the interface becomes RT unstable. Despite the appearance of these dark areas, the main edge of the cavity surface (near the centreline) can be identified by the edge tracking algorithm, as demonstrated by figure~\ref{fig:4}(b), and does not significantly affect the analysis.

The combustion of the gas mixture used to drive the implosion resulted in significant variation in the driving pressure applied to the annular piston over the timescale of the implosion. This driving pressure is an important parameter in determining the dynamics of the cavity implosion. Therefore, it was necessary to record the time evolution of the driving pressure to allow for a comparison of the experimental results to models. The pressure was measured using a PCB model 113A24 piezoelectric pressure sensor located 15~mm below the base of the piston. The pressure signals were recorded using an on-board oscilloscope at a rate of 40 MHz. An optical trigger indicator was used to synchronise the pressure sensor results from the rotating oscilloscope with the stationary high-speed cameras.

\section{Unperturbed motion of the cylindrically imploding rotating shell}
\label{sec:3.0}
\subsection{Model development}
\label{sec:3.05}

The evolution of perturbations on the surface of an imploding cylindrical shell is inherently coupled to the time-dependent evolution of the interface radius, velocity, and acceleration. This section will present a model for the unperturbed one-dimensional motion of the quasi-cylindrical imploding shell in the apparatus presented above under an external driving pressure and an internal cavity pressure. While the experimental arrangement used in this work resembles that of a cylindrical shell, it is complicated by the presence of curvature in the shell that transitions from axial motion to a cylindrically converging geometry as well as the presence of a heavy piston on the outer surface. An approach based on the conservation of energy during the implosion will be taken which can readily be extended to consider the unique geometry and the piston in a one-dimensional implosion model. The general model approach will be summarised in the text, while the complete model equations for a purely cylindrical geometry are presented in appendix~\ref{app:A}. The shell geometry and variables are shown schematically in figure~\ref{fig:3}. The subscript “$_0$” will be used to denote initial conditions. The model will consider the implosion of a shell composed of an incompressible and inviscid fluid initially in solid body rotation with no radial velocity. The gas within the cavity will be assumed to be calorically perfect with a constant specific heat ratio ($\kappa$) and will be compressed isentropically. The kinetic energy of the gas will be neglected due to its comparatively low density. The change in the surface energy of the cavity is also neglected due to the large radii (mm scale) considered in experiments. The conservation of energy for the system can be written as

\begin{equation}
\label{eq:3:1}
W_ {\mathrm{in}}=\Delta K_ {\mathrm{linear}}+\Delta K_ {\mathrm{rotational}}+\Delta I_ {\mathrm{gas}},
\end{equation}

\noindent where the work done on the system by the external pressure ($W_\mathrm{in}$) is equal to the sum of the change in the linear kinetic energy of the shell ($K_\mathrm{linear}$), the rotational kinetic energy of the shell ($K_\mathrm{rotational}$), and the internal energy of the gas within the cavity ($I_\mathrm{gas}$). Equation~\ref{eq:3:1} will be used to find an expression relating the radius ($R_1$) and radial velocity ($\dot{R}_1$) of the cavity surface. The motion of the cavity radius can thus be solved numerically by advancing the cavity radius and solving for the new radial velocity at each step. For an incompressible shell, the work done by the constant external pressure can be determined by

\begin{equation}
\label{eq:3:2}
W_ {\mathrm{in}}=\int_{V_{1_0}}^{V_1}P_2(t)\mathrm{d}V_1,
\end{equation}

\noindent where $V_1$ represents the volume of the cavity and $P_2$, the external pressure, can be either constant or time varying in which case the $W_\mathrm{in}$ term must be numerically integrated. The change in the cavity gas internal energy due to its isentropic compression can be expressed as

\begin{equation}
\label{eq:3:3}
\Delta I_ {\mathrm{gas}}=\frac{P_1V_1-P_{1_0}V_{1_0}}{\kappa-1},
\end{equation}

\noindent where $P_1$ and $\kappa$ are the pressure and specific heat ratio of the cavity gas, respectively. For a shell having no initial radial velocity, the change in the linear kinetic energy can be determined by evaluating the following integral:

\begin{equation}
\label{eq:3:4}
\Delta K_ {\mathrm{linear}}=\int_\mathrm{IS}^\mathrm{OS}\frac{1}{2}u^2 \mathrm{d}m,
\end{equation}

\noindent where $u$ is the linear velocity of the differential shell element d$m$. The integral is evaluated from the inner surface (IS) to the outer surface (OS) of the shell, including the driving piston. For an incompressible system, the volumetric flux is uniform across the shell, allowing for the linear velocity at any point to be expressed as a function of the radial velocity ($\dot{R}_1$) of the cavity surface:

\begin{equation}
\label{eq:3:5}
u=\frac{\dot{R}_1A_1}{A},
\end{equation}

\noindent where $A_1$ and $A$ are the area perpendicular to the flow velocity at the cavity surface and the fluid element d$m$ respectively. The change in the rotational kinetic energy can be expressed by the difference between the integrals for the initial and current rotational energies:

\begin{equation}
\label{eq:3:6}
\Delta K_ {\mathrm{rotational}}=\int_\mathrm{IS}^\mathrm{OS}\frac{1}{2}r^2\omega^2 \mathrm{d}m - \int_\mathrm{IS}^\mathrm{OS}\frac{1}{2}r_0^2\omega_0^2 \mathrm{d}m,
\end{equation}

\noindent where $\omega$ and $r$ are the radial position of the centroid and angular velocity of the differential shell element d$m$. The angular velocity of the element d$m$ can be determined from the initial rate of solid body rotation using the conservation of angular momentum:

\begin{equation}
\label{eq:3:7}
\omega=\omega_0\Big(\frac{r_0}{r}\Big)^2,
\end{equation}

\noindent where $r_0$ is the initial radial position of the centroid of the fluid element d$m$. The expressions for the change in the linear and rotational kinetic energies can thus be written as follows:

\begin{equation}
\label{eq:3:8}
\Delta K_ {\mathrm{linear}}=\int_\mathrm{IS}^\mathrm{OS}\frac{1}{2}\bigg(\frac{\dot{R}_1A_1}{A}\bigg)^2 \mathrm{d}m,
\end{equation}

\begin{equation}
\label{eq:3:9}
\Delta K_ {\mathrm{rotational}}=\int_\mathrm{IS}^\mathrm{OS}\frac{1}{2}\bigg(\frac{r_0^2\omega_0}{r}\bigg)^2 \mathrm{d}m - \int_\mathrm{IS}^\mathrm{OS}\frac{1}{2}r_0^2\omega_0^2 \mathrm{d}m.
\end{equation}

\noindent The integrals of equations~\ref{eq:3:8} and~\ref{eq:3:9} can readily be evaluated numerically for the experimental arrangement considered in this paper by discretising the shell into finite $\Delta m$ elements. As the implosion progresses, the change in position of the element boundaries can be determined from conservation of volume. As a result, all values in equation~\ref{eq:3:1} depend only on the position and velocity of the cavity surface, which gives an expression for the radial velocity of the cavity surface as a function of its radius. The motion of the cavity radius can thus be solved numerically by advancing the cavity radius and solving for the new radial velocity at each step. The time evolution and acceleration of the cavity surface are then determined by finite difference methods. The approach highlighted above is equivalent to that of applying the momentum equation for unsteady potential flow to the shell motion, which has been used extensively to study implosions in a purely cylindrical geometry \citep{Barcilon1974,Kull1991,Mikaelian2005}, but allows for simple treatment of the one-dimensional shell motion for the relatively complex geometry of the experimental arrangement.

The evolution in the centripetal acceleration of the liquid at the cavity surface must also be determined to evaluate the stability of the implosion. If it is again assumed that angular momentum is conserved, the centripetal acceleration ($a_\mathrm{c}$) at the shell inner surface can be determined as follows:

\begin{equation}
\label{eq:3:10}
a_\mathrm{c}=-\omega_0^2 \frac{R_{1_0}^4}{R_1^3}.
\end{equation}

\noindent The net acceleration ($a_\mathrm{net}$), which is expected to define the stability of the cavity, can then be determined by the sum of the radial and centripetal acceleration, defined as being positive for a net inward acceleration:

\begin{equation}
\label{eq:3:11}
a_\mathrm{net}=-(a_\mathrm{c}+\ddot{R}_1).
\end{equation}

\begin{figure}
  \centerline{\includegraphics[width=0.75\columnwidth]{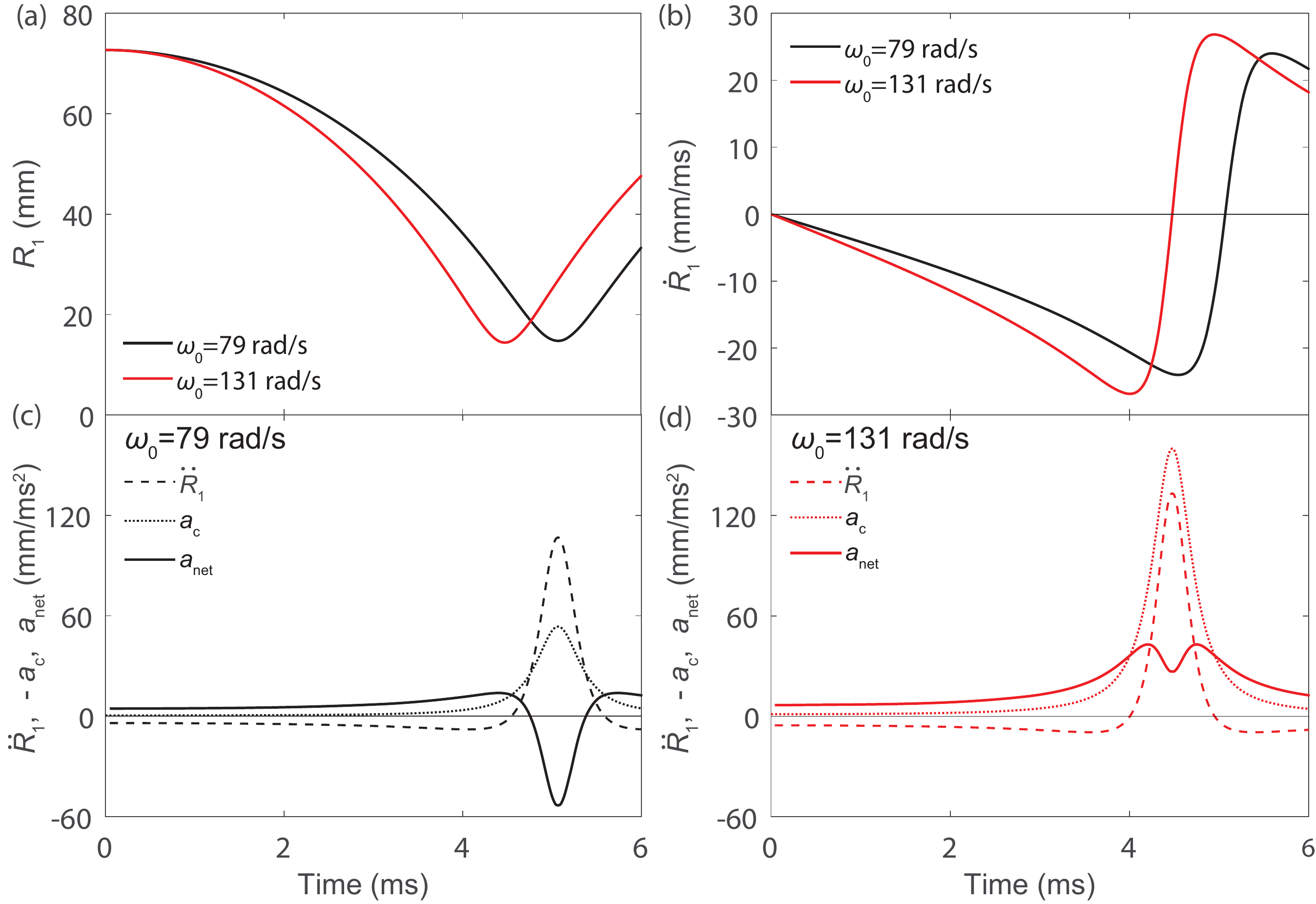}}
  \caption{Model results at two initial angular velocities showing the typical implosion dynamics of a rotating cylindrical shell: (a) shell inner radius as a function of time, (b) shell inner surface radial velocity as a function of time, (c) and (d) shell inner surface radial, centripetal, and net accelerations as a function of time for $\omega_0$ = 79~rad/s and 131~rad/s.}
  \label{fig:5}
\end{figure}

\noindent Figure~\ref{fig:5} presents the model predicted radius, velocity, and acceleration of the shell inner surface for conditions similar to those encountered in this paper. To demonstrate the effect of rotation, a relatively low initial angular velocity ($\omega_0$=79~rad/s) is compared to the case where the angular velocity is sufficient to maintain an inward-facing net acceleration throughout the implosion ($\omega_0$=131~rad/s). The simulations used the geometrical parameters highlighted in figure~\ref{fig:3} and a driving pressure ($P_2$) of 7.2 bar (79 rad/s) and 10.4~bar (131~rad/s). The internal cavity gas had an initial pressure of 1.2~bar and $\kappa$=1.18. Also shown in figure~\ref{fig:4} is the evolution in the liquid centripetal acceleration and the net acceleration at the cavity surface.

Figure~\ref{fig:5} demonstrates the expected dynamics of the imploding shell. Initially, the modest external pressure gradually accelerates the shell which accumulates kinetic energy. As the cavity approaches the point of maximum convergence and the shell surface reaches its maximum implosion velocity, a drastic increase in the cavity gas pressure results in the sudden deceleration of the shell inner surface. The focussing that results from the cylindrically imploding flow allows the cavity to reach pressures well above the external driving pressure. The radial deceleration, which reaches its peak at turnaround, is more than an order of magnitude greater than the maximum rate of acceleration during the implosion phase. At this stage, conservation of angular momentum leads to a significant increase in the angular velocity and therefore of the centripetal acceleration of the liquid at the cavity surface. As can be seen in the acceleration graphs, for the slow rotating case, the outward radial acceleration grows faster than the centripetal acceleration leading to a net outward acceleration at the cavity surface, while for the fast-rotating case, the net acceleration remains positive.

\subsection{Parametrisation of the experimental design space}
\label{sec:3.1}

This section will present the methodology with which the model for the unperturbed one-dimensional motion of the inner surface of the liquid shell was used to select the initial rotation rates and driving conditions used in experiments. The goal was to find a set of parameters which could adequately demonstrate the effect of varying the inner surface centripetal acceleration on the stability of the shell inner surface during the implosion. The parameterization presented below also allows the results of this study to be generalised and applied to other experimental conditions. For the purpose of the parameterization, it will be assumed that the driving pressure ($P_2$) remains constant throughout the implosion. The time evolution of the imploding shell inner surface ($R_1$($t$)) is thus a function of seven experimental initial conditions:

\begin{equation}
\label{eq:3:12}
R_1=f(t;P_{1_0},P_2,\kappa,R_{1_0},R_2,\omega_0,\rho),
\end{equation}

\noindent With such a large number of parameters, it is not trivial to choose experimental settings that allow for a wide variation in the magnitude of the centripetal acceleration near turnaround, while producing implosions that can be directly compared to one another by maintaining the same level of convergence ($R_{1_0}/R_{1_{\mathrm{min}}}$) and target compression ($P_{1_{\mathrm{max}}}/P_{1_0}$). Additional constraints are placed by the fact that the apparatus has a number of practical limitations which must be considered, including the maximum angular velocity of the apparatus ($\omega_{0_{\mathrm{max}}}\approx 131$~rad/s), the specific heat ratio of common gases ($\kappa\approx 1.1$--1.67), minimum combustion pressure for mixture ignition ($P_{2_{\mathrm{min}}}\approx 7$~bar), and the dimensions of the apparatus ($R_{1_0}<~$75~mm and $R_2= 187$~mm).

The seven variables of equation~\ref{eq:3:12} can be grouped into the following four non-dimensional parameters.
\refstepcounter{equation}
$$
\frac{R_2}{R_{1_0}}, \quad
\frac{P_2}{P_{1_0}}, \quad
\kappa, \quad
\frac{\rho \omega_0^2 R_{1_0}^2}{P_2},
\eqno{(\theequation{\mathit{a},\mathit{b},\mathit{c},\mathit{d}})}
\label{eq:3:13}
$$

These represent the initial shell thickness ratio, the initial pressure ratio between the inner and outer surface, the specific heat ratio of the cavity gas, and the ratio of the initial dynamic pressure from the rotation of the shell to the driving pressure, respectively. These scale-independent parameters are the non-dimensional initial conditions which determine the dynamics of the imploding shell, including the evolution of the radial and centripetal acceleration on the cavity surface. For this paper, model parameters were scaled by their initial conditions rather than the common practice of non-dimensionalising using values at maximum convergence \citep{Barcilon1974}, because the goal was to determine appropriate experimental initial conditions which would produce rotationally stabilised implosions.

In comparing implosions, it is convenient to summarise the influence of rotation on the implosion by the ratio of the centripetal acceleration to radial acceleration at turnaround ($-a_{\mathrm{c}_\mathrm{max}}/\ddot{R}_{1_\mathrm{max}}$). A ratio much smaller than unity indicates that the centripetal acceleration is negligible compared to the radial acceleration near turnaround and is not expected to contribute significantly to the net surface acceleration and the stability of the interface. As the ratio increases towards unity, the centripetal acceleration has an increasingly large contribution to the net acceleration near turnaround, with a ratio that is greater than unity indicating that the net acceleration remains inward-directed throughout the implosion, possibly stabilising the surface from RT perturbation growth.

\begin{figure}
\centerline{\includegraphics[width=0.7\columnwidth]{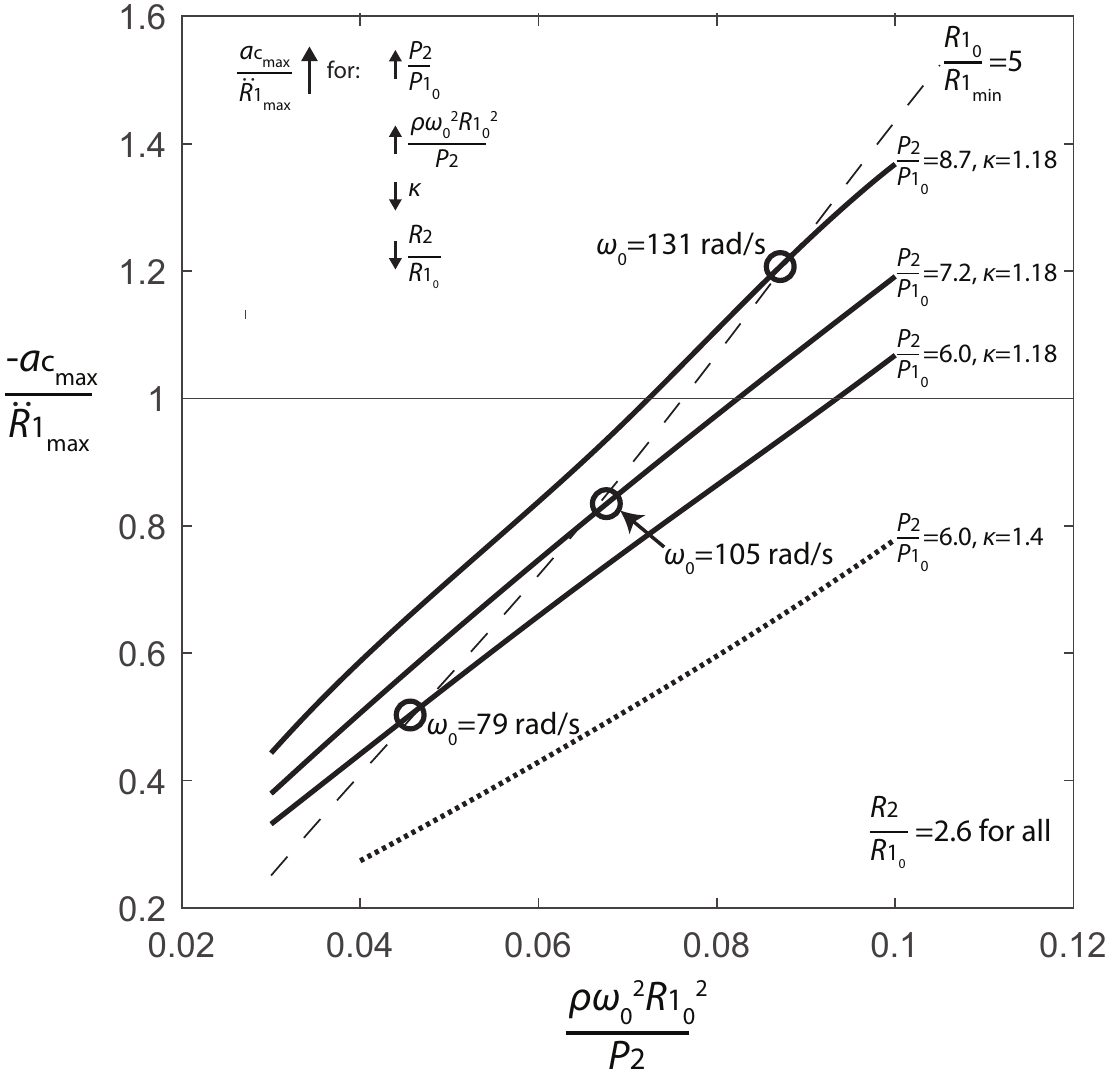}}
\caption{Parameter map showing the effect of the initial rotation rate and driving conditions on the ratio of centripetal acceleration to radial acceleration at the cavity inner surface at turnaround.}
\label{fig:6}
\end{figure}

The effect of the four non-dimensional parameters on the ratio of the centripetal and radial acceleration at turnaround is summarised in figure~\ref{fig:6}, which plots the ratio of maximum centripetal acceleration to radial acceleration as a function of the rotational-pressure-based non-dimensional term. The data was obtained using the energy-based implosion model presented in section~\ref{sec:3.05}. The solid lines represent curves where the thickness ratio, pressure ratio, and specific heat ratio are held constant to isolate the effect of rotation. As expected, an increase in the ratio of dynamic rotational pressure to driving pressure results in an increase in the influence of the centripetal acceleration at turnaround. It should be noted that the convergence ratio of the implosion ($R_{1_0}/R_{1_{\mathrm{min}}}$) is reduced as the initial angular velocity is increased due to the greater centrifugal force of the fluid. The dashed line is a contour of implosions having a convergence ratio of $R_{1_0}/R_{1_\mathrm{min}}$=5, which provides a measure of the increase in $P_2$ that is required to reach the same final implosion radius as $\omega_0$ is increased. The ratio of centripetal to radial acceleration at turnaround can also be increased by lowering $\kappa$, $R_2/R_{1_0}$, or $P_{1_0}$, which provides additional convergence, thus increasing the angular velocity of the fluid at turnaround due to angular momentum conservation.

The non-dimensional parameter analysis presented above is expected to apply to a wide range of experiment length and time scales, the limitation being the assumption of conservation of angular momentum within the shell, which should be valid as long as the Reynolds number of the flow is much greater than unity, ensuring that the inertial forces far outweigh the viscous forces. For the conditions probed in this work, this corresponds to a requirement that the shell initial radius be on the order of 1~mm or greater.

The parameter map of figure~\ref{fig:6} was used to select the initial rotation rates and driving conditions which were used to experimentally verify the effect of rotation on the stability of imploding cavities. The three chosen settings, shown by circular markers and labelled by their initial angular velocity in figure~\ref{fig:6}, correspond to a variation in initial angular velocity from a case where the maximum centripetal acceleration is only a small portion of the peak radial acceleration (79~rad/s), up to a case where the centripetal acceleration is sufficiently large to allow the net acceleration to remain inward-directed for the entire implosion (131~rad/s). All settings used the same shell thickness ratio, cavity gas specific heat ratio, and initial cavity gas pressure. The points all lie on the contour corresponding to a convergence ratio of~5, thus resulting in geometrically similar implosions which can directly be compared to each other. This was achieved by increasing the outer driving pressure ($P_2$) in proportion to the initial angular velocity of the shell. Ethane was used as the cavity gas due to the fact that its low specific heat ratio ($\kappa$=1.18) relaxes the need for high initial angular velocities ($\omega_0$) or low initial cavity fill pressures ($P_{1_0}$) to produce implosions that have a net inward acceleration at turnaround. The model predictions for the 79~rad/s and 131~rad/s experiments presented in figure~\ref{fig:5} correspond to the chosen experimental parameters. As can be seen, the implosion results in similar radial accelerations and implosion timescales, allowing for a direct comparison of perturbation growth between experiments.

\subsection{Model verification}
\label{sec:3.2}

\begin{figure}
\centerline{\includegraphics[width=0.5\columnwidth]{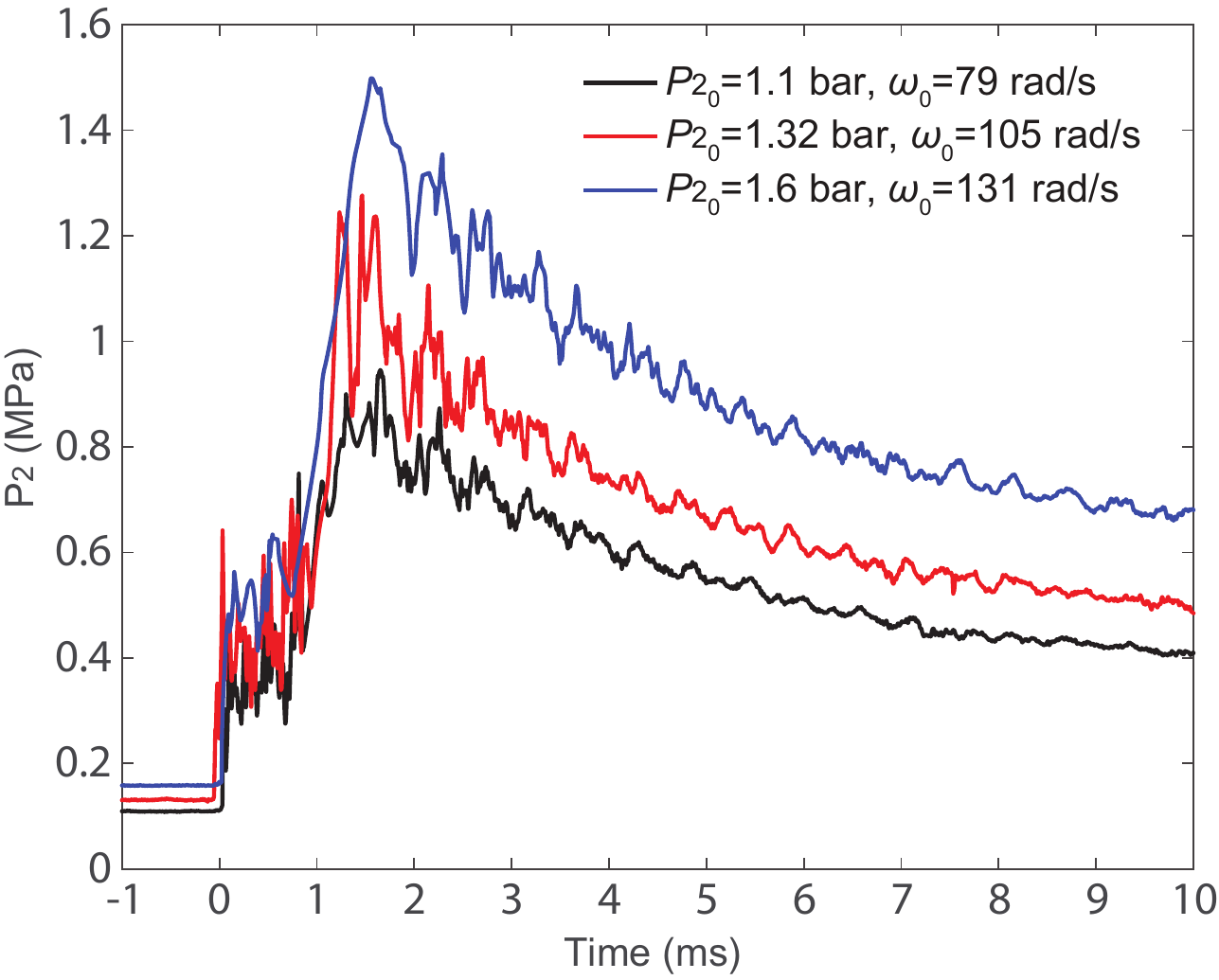}}
\caption{Measured time evolution of the driving pressure ($P_2$) for the model verification experiments.}
\label{fig:7}
\end{figure}

\begin{figure}
\centerline{\includegraphics[width=1.0\columnwidth]{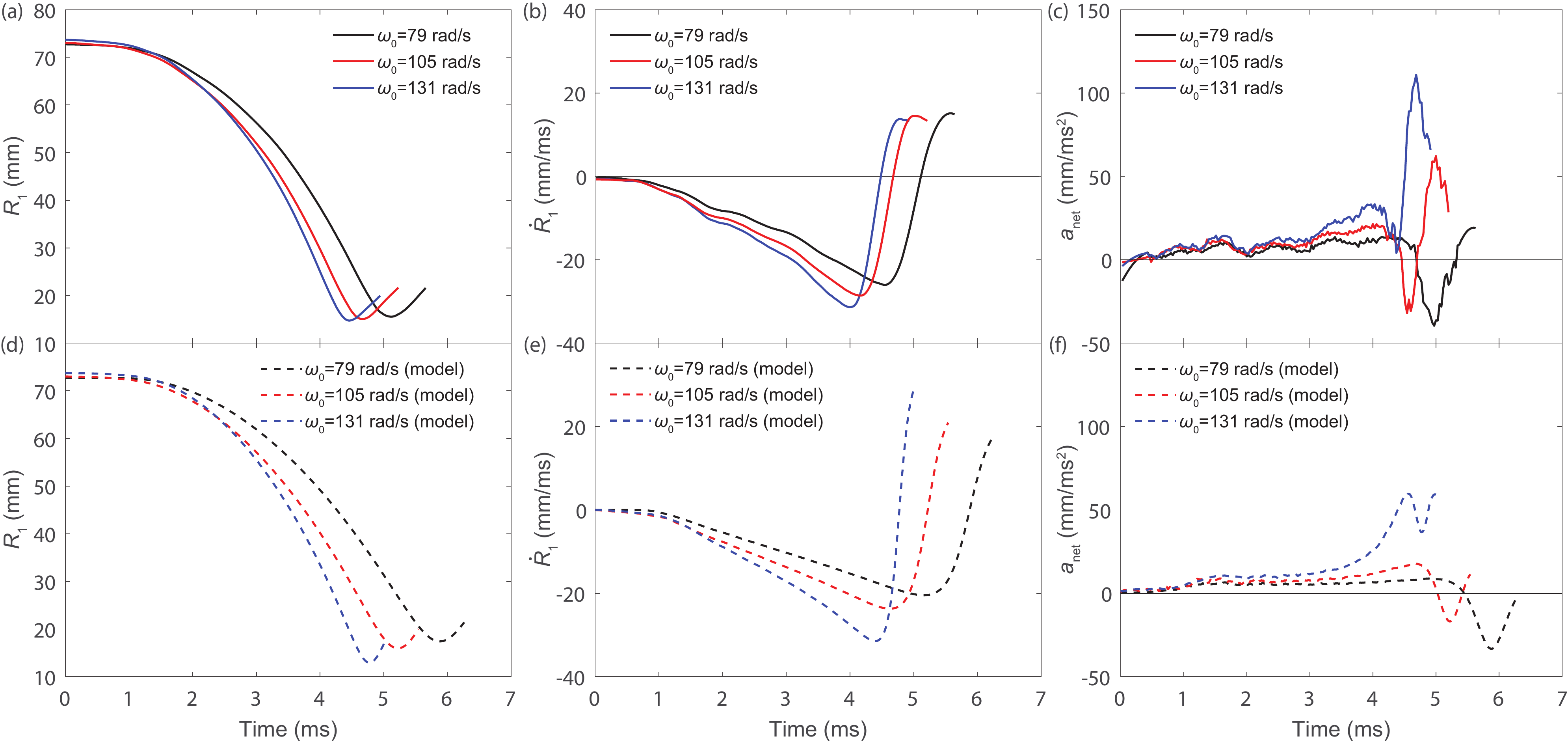}}
\caption{Time evolution of the shell inner surface radius, velocity, and net acceleration for (a)-(c) the model calibration experiments, and (d)-(f) the corresponding model results.}
\label{fig:8}
\end{figure}

Initial experiments were performed to verify the validity of the energy-based model described in section~\ref{sec:3.05} and ensure that the chosen experiment initial rotation rates and driving conditions highlighted in figure~\ref{fig:6} resulted in the desired implosion properties. The initial angular velocities of 79, 105, and 131~rad/s had corresponding initial combustible gas fill pressures ($P_{2_0}$) of 1.1, 1.32, and 1.6~bar, respectively, in order to reach similar convergence ratios. A summary of the experimental parameters can be seen in table~\ref{tab:1}. The time evolution of the driver gas pressure was recorded for the experiments and is plotted in figure~\ref{fig:7}. As can be seen, the combustion process causes the pressure to ramp up over approximately 1.5~ms as combustion occurs and is followed by a gradual pressure decay as the gasses cool, leading to a significant variation in the driving pressure over the timescale of the experiment ($\approx$6~ms). Varying the initial gas pressure resulted in the desired proportional increase in the driving pressure. The evolution in the radius, velocity, and net acceleration of the inner cavity surface for the experiments is presented in figure~\ref{fig:8}. The radius data was collected with a high-speed camera at a frame rate of 35 kHz and a spatial resolution of 0.4~mm/pixel, while the velocity and acceleration were calculated by differentiating the radius-time data. The centripetal acceleration at the shell inner surface was calculated from the measured initial and current radius assuming conservation of angular momentum. Also shown in figure~\ref{fig:8} are the curves for the model of section~\ref{sec:3.05} for the corresponding experiments. The curves were generated using the pressure data plotted in figure~\ref{fig:7} as an input.

\begin{table}
\begin{center}
\begin{tabular}{ccccccc}
$\omega_0$ & $P_2$ fill & $P_{1_0}$ & $\kappa$ (gas) & $R_{1_0}\textsuperscript{a}$ & $R_2$ & $\rho$ \\
(rad/s) & (bar) & (bar) & & (mm) & (mm) & ($\mathrm{g/cm^3}$)\\
\hline\noalign{\smallskip}
79 & 1.1 & 1.2 & 1.18 (ethane) & 74 & 187 & 1.0 \\
105 & 1.32 & 1.2 & 1.18 (ethane) & 74 & 187 & 1.0 \\
131 & 1.6 & 1.2 & 1.18 (ethane) & 74 & 187 & 1.0 \\
\multicolumn{7}{c}{\textsuperscript{a}nominal values, $\approx$ 1~mm variation existed between experiments}\\
\end{tabular}
\caption{Initial rotation rates and driving conditions used in experiments.}
{\label{tab:1}}
\end{center}
\end{table}

\begin{figure}
\centerline{\includegraphics[width=1.0\columnwidth]{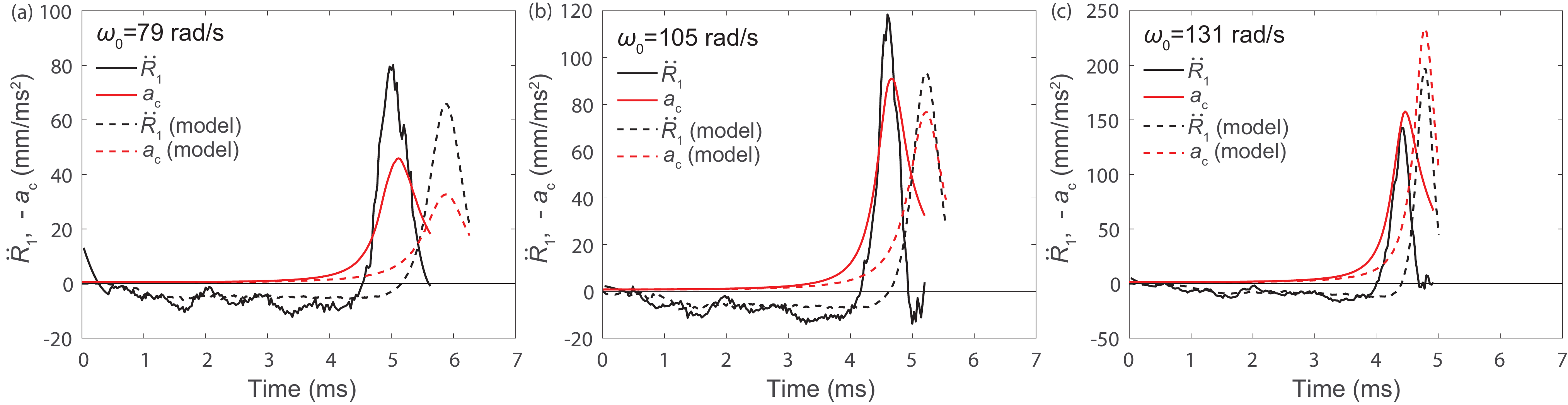}}
\caption{Radial and centripetal acceleration of inner surface of liquid shell as a function of time for the model verification experiments and the corresponding model results at $\omega_0$= (a) 79, (b) 105, and (c) 131~rad/s.}
\label{fig:9}
\end{figure}

The experimental data agrees relatively well with the model, notably with respect to the radius at full convergence, maximum implosion velocity, and the general character and minimum values of the net acceleration. There is some discrepancy between the model and experiments with regards to the timescale over which the implosion occurs, but this does not materially change the dynamics of the cavity acceleration near turnaround. The breakdown between the radial and centripetal accelerations for the three experiments and the corresponding model runs are shown in figure~\ref{fig:9}. The experimentally observed acceleration profiles are similar to the model for all three cases, and as expected, there is a large increase in both the radial and centripetal acceleration as the turnaround radius is approached. For the low initial angular velocity experiments, the rate at which the radial acceleration increases near turnaround exceeds that of the centripetal acceleration, causing the surface to have a net outward acceleration near the minimum radius, while for the 131~rad/s case, the angular momentum of the surface is sufficiently large to maintain a net inward-facing acceleration. It should be noted that the rate at which the radial acceleration decays after reaching its maximum value at turnaround is notably greater in experiments than the model. This is likely caused by the fact that edge effects cause drops and spray to remain on the windows, making it difficult to capture the true radial position of the cavity surface beyond turnaround.

Focussing on the experimental data, the goal of choosing initial rotation rates and driving conditions which create implosions that have similar dynamics in terms of timescale, convergence ratio, and peak radial acceleration, but significantly different net acceleration profiles was largely achieved. The radius at turnaround decreased slightly with increasing angular velocity due to limitations in predicting the correct initial fill pressure, but the convergence ratios of the three experiments were within 7\% of each other. The increase in convergence ratio also resulted in greater peak radial accelerations for the high initial angular velocity experiments, which nonetheless were quite similar for all experiments. Most importantly, the variation in the initial angular velocity resulted in three significantly different net acceleration profiles near turnaround: a large outward net acceleration (79~rad/s), a small outward net acceleration (105~rad/s), and an inward acceleration for the entire implosion (131~rad/s). 

\section{Experimental results}
\label{sec:5.0}

\subsection{Unperturbed experiments}
\label{sec:5.1}
\begin{figure}
	\centering
	\includegraphics[width=0.85\columnwidth]{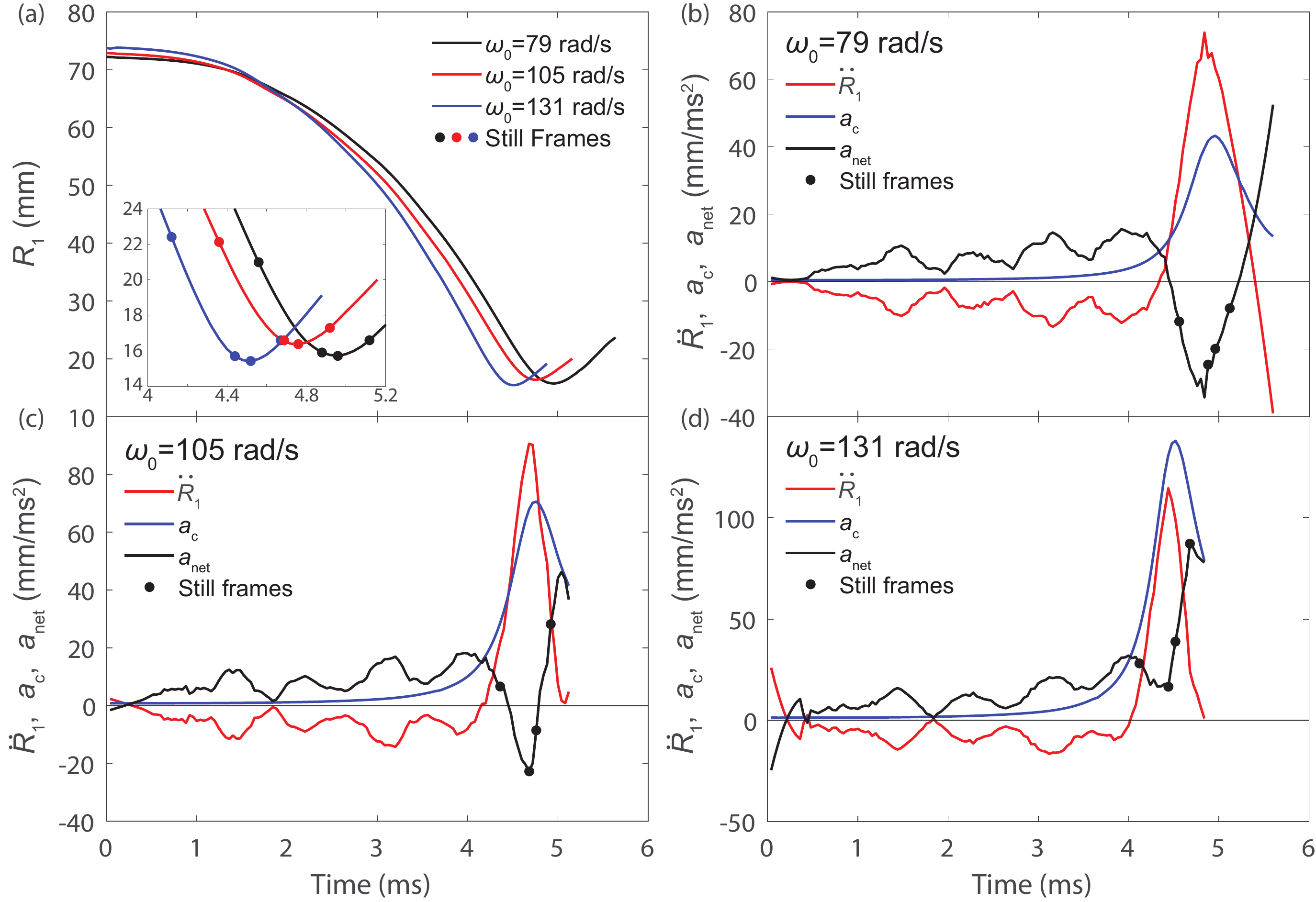}
	\caption{(a) measured time evolution of the inner radius of the liquid shell for the unperturbed experiments, with the circular markers indicating the time corresponding to the images in figure~\ref{fig:11}. (b)-(d) measured radial, centripetal, and net acceleration of the liquid shell inner surface for the $\omega_0$=(b) 79, (c) 105, and (d) 131~rad/s experiments. }
	\label{fig:10}
\end{figure}

\begin{figure}
	\centering
	\includegraphics[width=0.5\columnwidth]{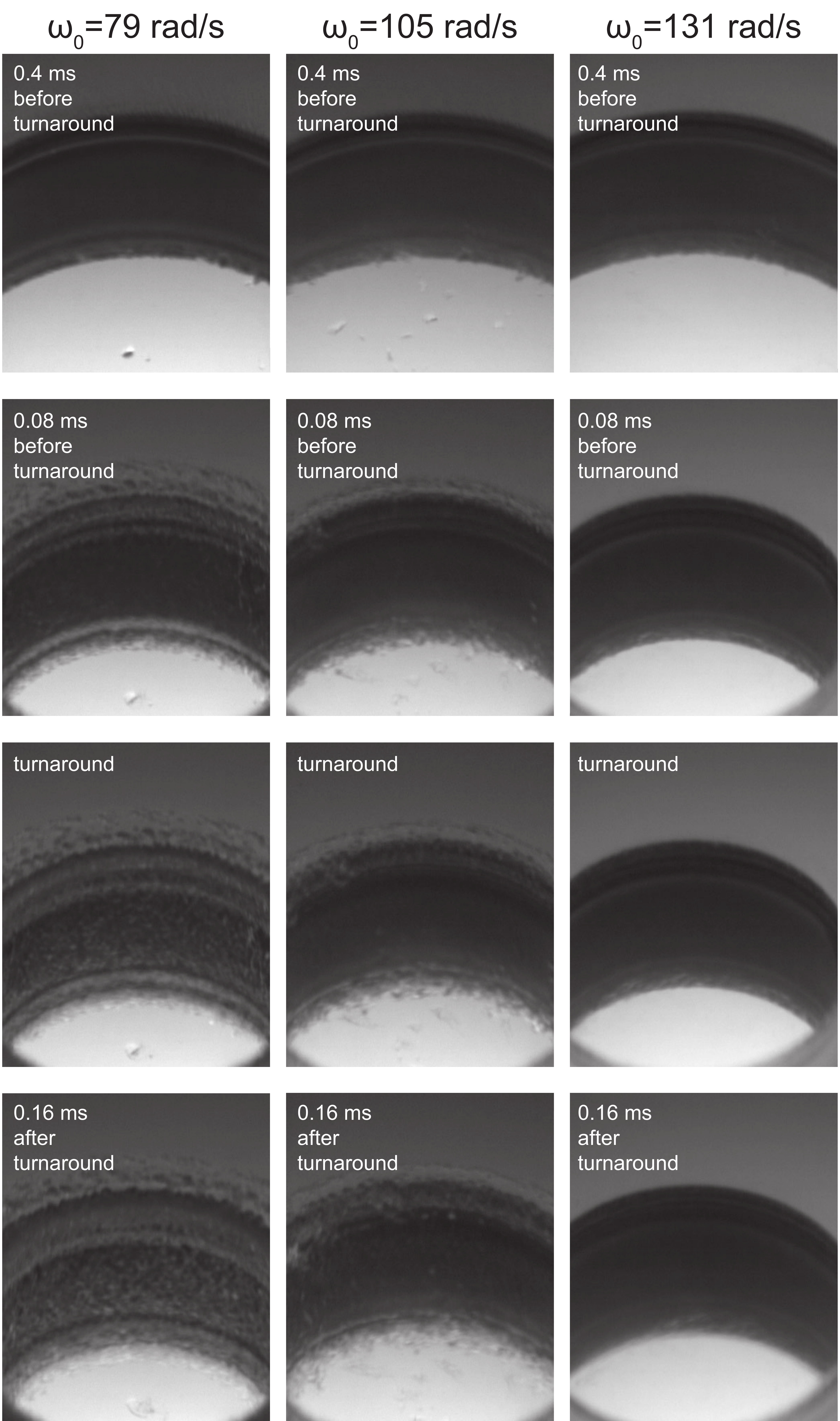}
	\caption{Off-axis view images of the inner surface of the liquid shell for the unperturbed experiments at progressive stages of cavity collapse.}
	\label{fig:11}
\end{figure}

The implosion of nominally smooth cavities was used to study the effect of shell rotation on the growth of spray-like high-mode-number perturbations that are prone to form as the gas-liquid interface becomes RT unstable. Two high-speed cameras operating at a synchronised frame rate of 25~kHz were used to simultaneously monitor the evolution in the shell inner radius with a normal view (spatial resolution of approximately 0.3~mm/pixel) and the condition of the cavity surface with an off-axis view. Three experiments were performed using the rotation rates and driving conditions described in table~\ref{tab:1}. Figure~\ref{fig:10} shows the evolution in the cavity radius and acceleration for all three experiments, while figure~\ref{fig:11} shows a series of selected off-axis images taken at equivalent implosion times relative to turnaround. Videos showing off-axis and on-axis views of the complete implosion can be seen in movies~1 and~2 of the supplementary material. The time corresponding to the images of figure~\ref{fig:11} is marked by circular markers on the radius-time and acceleration-time plots of figure~\ref{fig:10}. The plots of the inner surface radius and acceleration show that all three implosions again reached similar maximum convergence ratios and had notably different net acceleration profiles. The first row of off-axis images shows that the cavity surface had a mirror-like appearance up to 0.4~ms prior to turnaround for all three experiments. Bands can be seen at the top and bottom edges of the cavity due to boundary layer growth on the windows of the test section. These bands broadened as the implosion proceeded, particularly for the 79~rad/s experiment, but did not appear to affect the central portion of the cavity surface. As the cavity converged further, signs of perturbation growth began to form on the surface of the low angular velocity ($\omega_0$=79~rad/s) cavity, which developed into a spray-like foam at later times. By correlating the off-axis images to the circular markers on the acceleration-time plot of figure~\ref{fig:10}, it can be seen that the surface perturbations were only observed after the net acceleration of the shell inner surface was directed outward. The spatial and time resolution of the videos did not allow for measurements of the amplitude and growth rate of the high wavenumber perturbations. At the intermediate initial angular velocity ($\omega_0$=105~rad/s), the growth of spray-like perturbations was delayed until turnaround, where small amplitude perturbations appeared as white marks on the previously uniform dark surface. In the case of the highest initial angular velocity ($\omega_0$=131~rad/s), the growth of high-mode-number perturbations was completely suppressed, with the inner surface maintaining a mirror-like appearance beyond turnaround.

\subsection{Low-mode-number experiments}
\label{sec:5.2}

Experiments with a mode-6 obstruction plate were performed to obtain perturbation growth measurements that could be compared with the model developed in section~\ref{sec:4.0}. The comparatively large size and lower growth rates of the mode-6 perturbations allowed for reliable measurements of perturbation amplitudes as a function of time, which was not possible for the high wavenumber spray observed in the experiments presented in section~\ref{sec:5.1}. The mode-6 perturbations were seeded by adding the obstruction plate pictured in figure~\ref{fig:2} to the test section. The plate was composed of six rounded fins with a tip diameter of 5~mm, which were located 9.5~mm behind the initial position of the shell inner surface. The initially smooth cavity surface was perturbed during the early stages of the implosion due to the drag induced by the obstruction plate fins, which caused a velocity deficit at the portion of the shell inner surface near the tips (see movie~3 of the supplementary material). As the cavity imploded further, the influence of the fins on the inner surface disturbances was assumed to be negligible, with the evolution of the perturbation being dominated by the convergence and instability mechanisms described in section~\ref{sec:1.0}. The rotation rates and driving conditions described in table~\ref{tab:1} were once again used for these experiments. To resolve the perturbation growth near turnaround with sufficient spatial resolution, two experiments were performed at each angular velocity, one with a normal view of the entire cavity to track the early evolution of the perturbations, the other with a zoomed-in normal view to capture the perturbation growth near turnaround along with a synchronised off-axis view to image the cavity surface in a similar manner to the unperturbed experiments. The full-view experiments had a spatial resolution of approximately 0.31~mm/pixel at a frame rate of 35~kHz, while the zoomed-in experiments had a spatial resolution of 0.14~mm/pixel at a frame rate of 25~kHz.

\begin{figure}
\centering
\includegraphics[width=1.0\columnwidth]{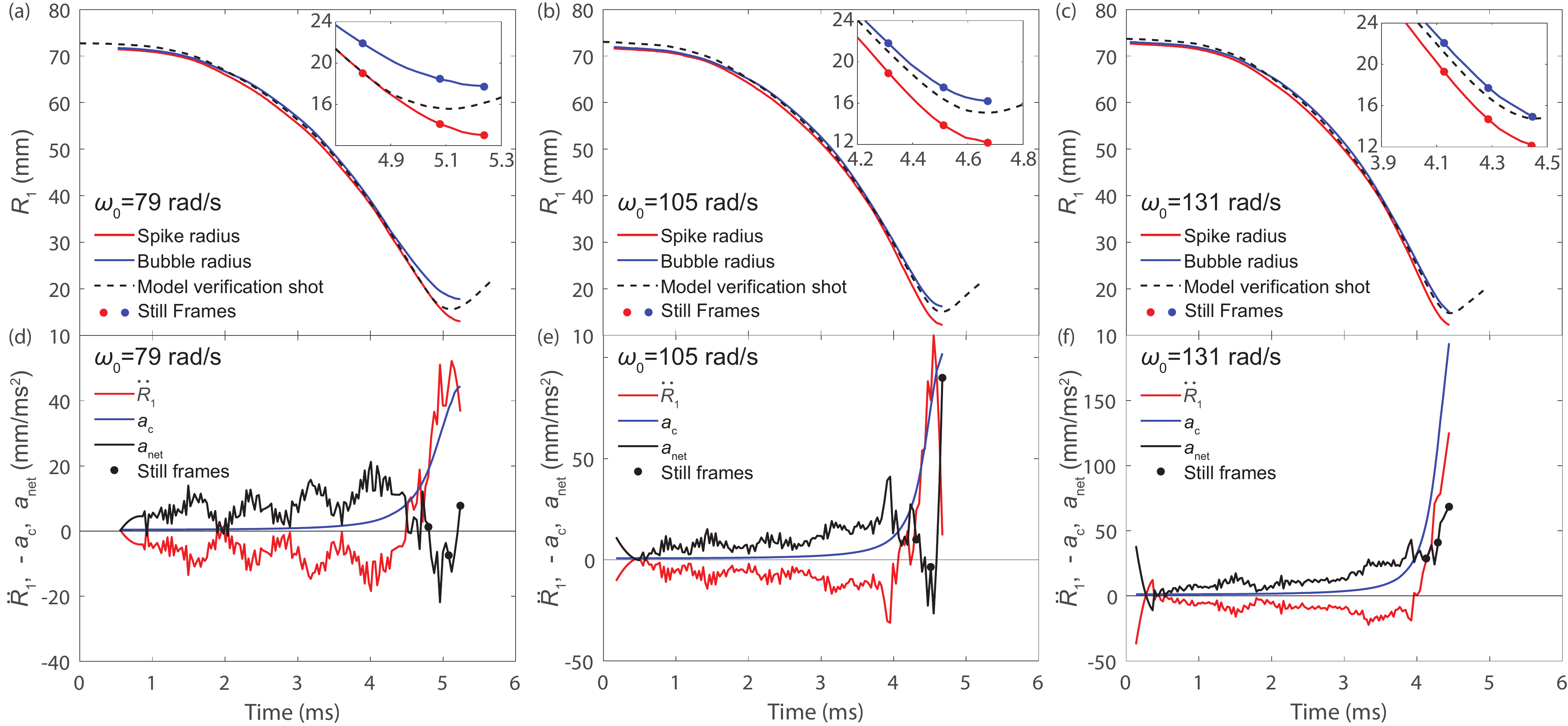}
\caption{(a)-(c) measured time evolution of the perturbation spike and bubble radius of the perturbed mode-6 experiments for $\omega_0$=(a) 79, (b) 105, and (c) 131~rad/s. The shell inner radius for the model verification experiments of figure~\ref{fig:8}(a) are also shown for reference. (d)-(f) average liquid shell inner surface radial, centripetal, and net acceleration for the $\omega_0$=(d) 79, (e) 105, and (f) 131~rad/s experiments. The circular markers indicate the time corresponding to the images in figure~\ref{fig:13}.}
\label{fig:12}
\end{figure}

\begin{figure}
\centering
\includegraphics[width=0.75\columnwidth]{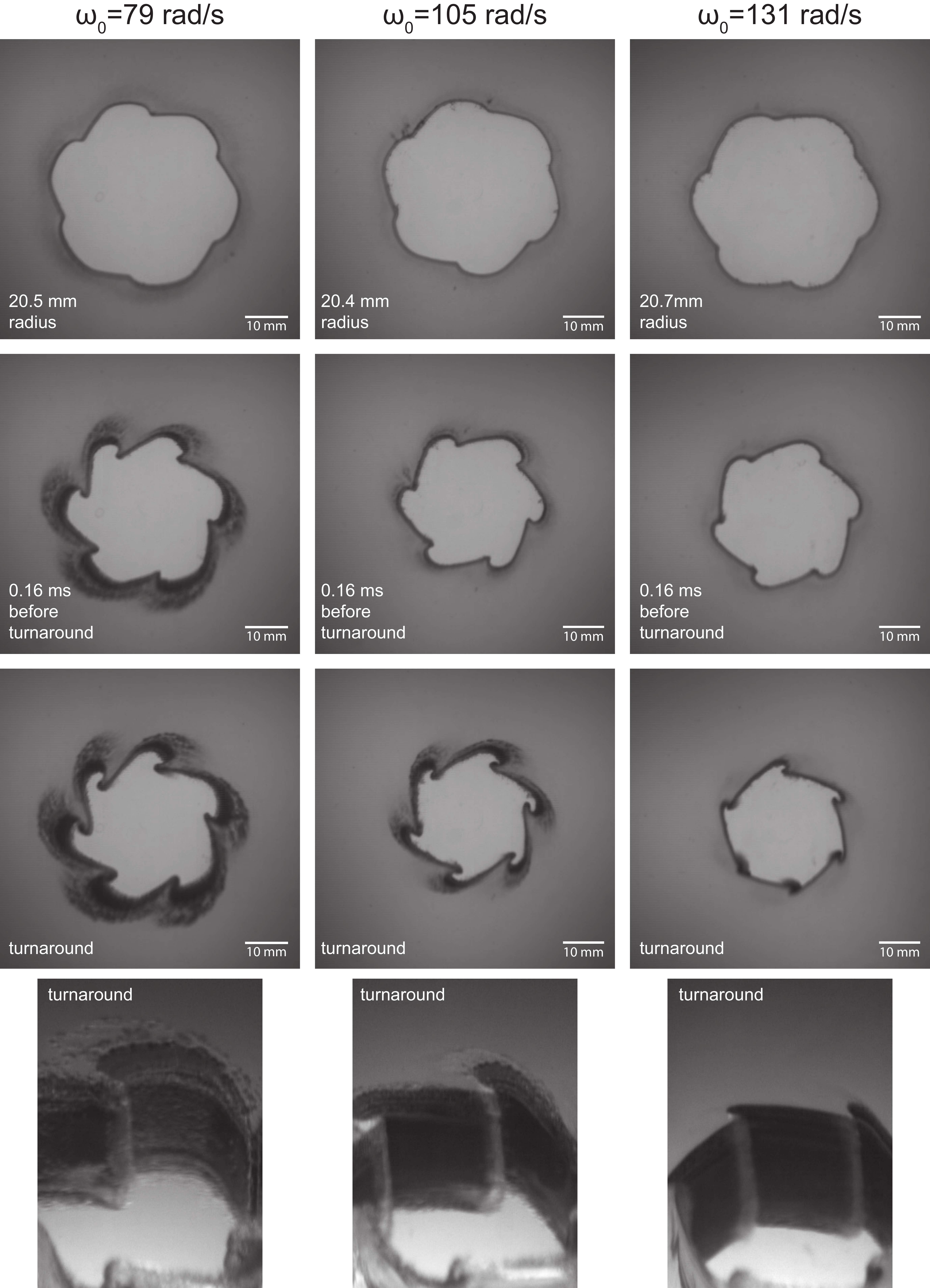}
\caption{Images of the mode-6 perturbed experiments for the three initial rotation rates including a normal view of the cavity at various stages of collapse and an off-axis view image at turnaround.}
\label{fig:13}
\end{figure}

The time evolution of the spike and bubble radius, as defined by figure~\ref{fig:4}, is shown in figure~\ref{fig:12}(a)-(c). For comparison, the radius vs time data for the unperturbed model verification experiments presented in figure~\ref{fig:8}(a) are also included. The shell inner surface acceleration, calculated from numerical differentiation of the average of the spike and bubble radius, is presented in figure~\ref{fig:12}(d)-(f). Key images corresponding to the three implosion conditions are presented in figure~\ref{fig:13}. The first three rows of images in figure~\ref{fig:13} are normal views of the cavity at sequential times, while the final image is an off-axis view of the cavity surface at turnaround. Full videos of the normal view (movie~3 and movie~4) and off-axis view (movie~5) for the three initial rotation rates can be seen in the supplementary material. As can be seen from the first row of images, the perturbation growth at a radius of approximately 21~mm was nearly identical for all three experiments. The following two sets of images correspond to 0.16~ms prior to reaching full convergence and turnaround, respectively. At this point in the implosion, a significant reduction in the growth rate of the perturbations was observed with increasing initial angular velocity. The circular markers of figure~\ref{fig:12}, which indicate the time location of the images relative to the evolution in the cavity acceleration, show that the significant difference in perturbation growth seen in the images occurred after the net surface acceleration became outward facing in the 79~rad/s and 105~rad/s experiments. Both the on-axis and off-axis images of figure~\ref{fig:13} show that the perturbations became increasingly asymmetric as the implosion proceeded in all three cases, eventually taking on the shape of a breaking wave near turnaround. The off-axis images also show the simultaneous growth of high-mode-number spray-like perturbations on the cavity surface for the 79~rad/s and 105~rad/s experiments. The boundary layer at the top and bottom edges of the cavity can be seen on the off-axis view as bands, and lead to a broadening of the cavity edges in the on-axis images. In the 79~rad/s and 105~rad/s experiments, off-axis images show flow disturbances near the perturbation bubbles along the top window surface, which caused the formation of the slightly darkened areas away from the cavity edge in the on-axis images (see labelled effect in figure~\ref{fig:4}(b)). Despite these edge effects, the perturbations could be reliably tracked up to turnaround in all cases.
 
\begin{figure}
	\centering
	\includegraphics[width=1.0\columnwidth]{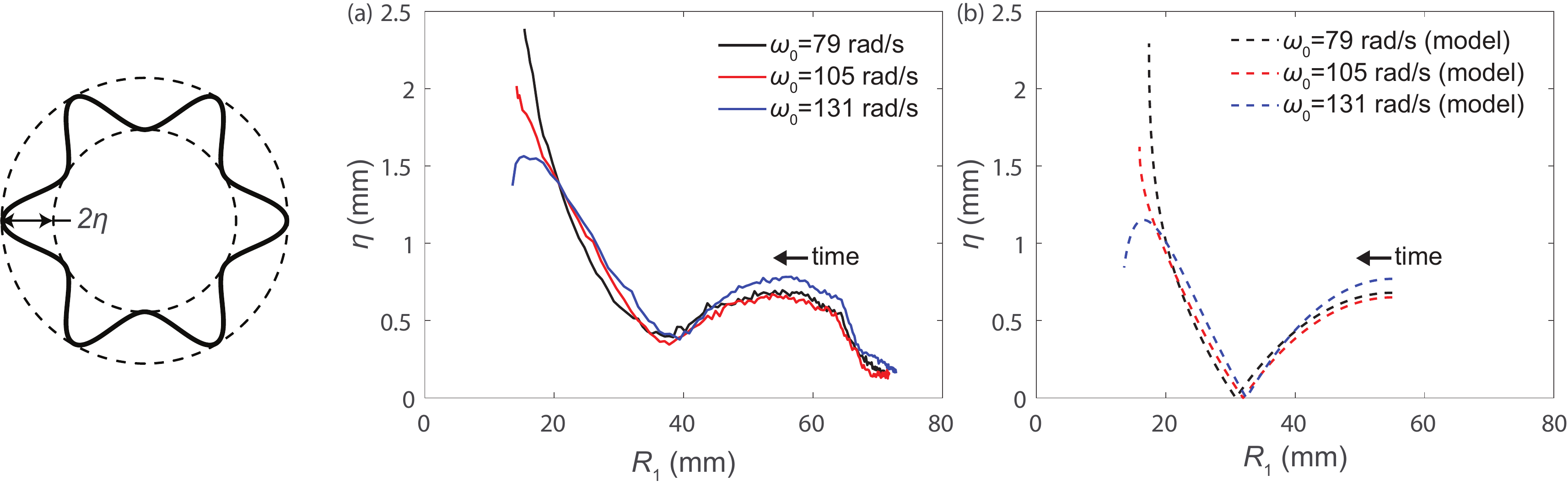}
	\caption{Perturbation amplitude as a function of radius for (a) the mode-6 number experiments and (b) the corresponding model results from equation~\ref{eq:4:20} using the model data from figure~\ref{fig:8}(d)-(f) as inputs.}
	\label{fig:14}
\end{figure}

The evolution in the amplitude of the low-mode-number perturbations as a function of the radius for the three initial rotation rates is plotted in figure~\ref{fig:14}(a). The implosion had three main phases: the initial growth in the perturbations caused by the obstacle plate, oscillation of the perturbations including a phase reversal at a radius of approximately 35~mm, and the final phase where the perturbation amplitude growth was sensitive to the cavity surface dynamics near turnaround. From figure~\ref{fig:14}(a), it is evident that the perturbation amplitude was nearly identical at a radius of 21~mm which corresponds to the first set of images in figure~\ref{fig:13}, but diverged significantly beyond this point. Near turnaround, the perturbations grew at an increasing rate for the 79~rad/s experiment, while the growth occurred at a nearly constant rate in the 105~rad/s experiment. For the 131~rad/s experiment, the perturbation growth stalled and reversed as turnaround was approached, which is indicative of a phase reversal. Figure~\ref{fig:14}(b) shows the perturbation growth predicted by equation~\ref{eq:4:20}, using the radius, velocity, and radial acceleration from the simulations presented in figure~\ref{fig:8}(d)-(f) as inputs. The model was initialised by taking the experimentally measured perturbation amplitude at the beginning of the phase reversal, where the perturbation amplitude plateaus and the growth rate ($\dot{\eta}$) goes to zero. This corresponded to a cavity radius of 55~mm and initial perturbation amplitudes of 0.68, 0.65, and 0.77~mm for the 79, 105, and 131~rad/s experiments respectively. By applying equation~\ref{eq:4:20}, it is inherently assumed that the fins, which are 5.7 tip diameters away from the cavity surface when the model is initialised, have a negligible effect on subsequent perturbation growth. As can be seen in figure~\ref{fig:14}, both the magnitude and the general characteristics of the evolution in the perturbation amplitude predicted by the model were similar to those seen in experiments. The oscillation of the perturbations in the early implosion was captured by the model and was again seen to be independent of the initial angular velocity. More importantly, the notable effect of initial angular velocity on the perturbation growth near turnaround was captured by the model, which predicted an increasing rate of perturbation growth for the two low angular velocity experiments as well as the experimentally observed decay in perturbation amplitude for the 131 rad/s case. The only notable discrepancy between the model and the experiment occurs at turnaround for the 79~rad/s and 105~rad/s experiments, where the growth rate of the perturbations was over-predicted by the model.

\section{Discussion}
\label{sec:6.0}

\begin{figure}
\centering
\includegraphics[width=0.8\columnwidth]{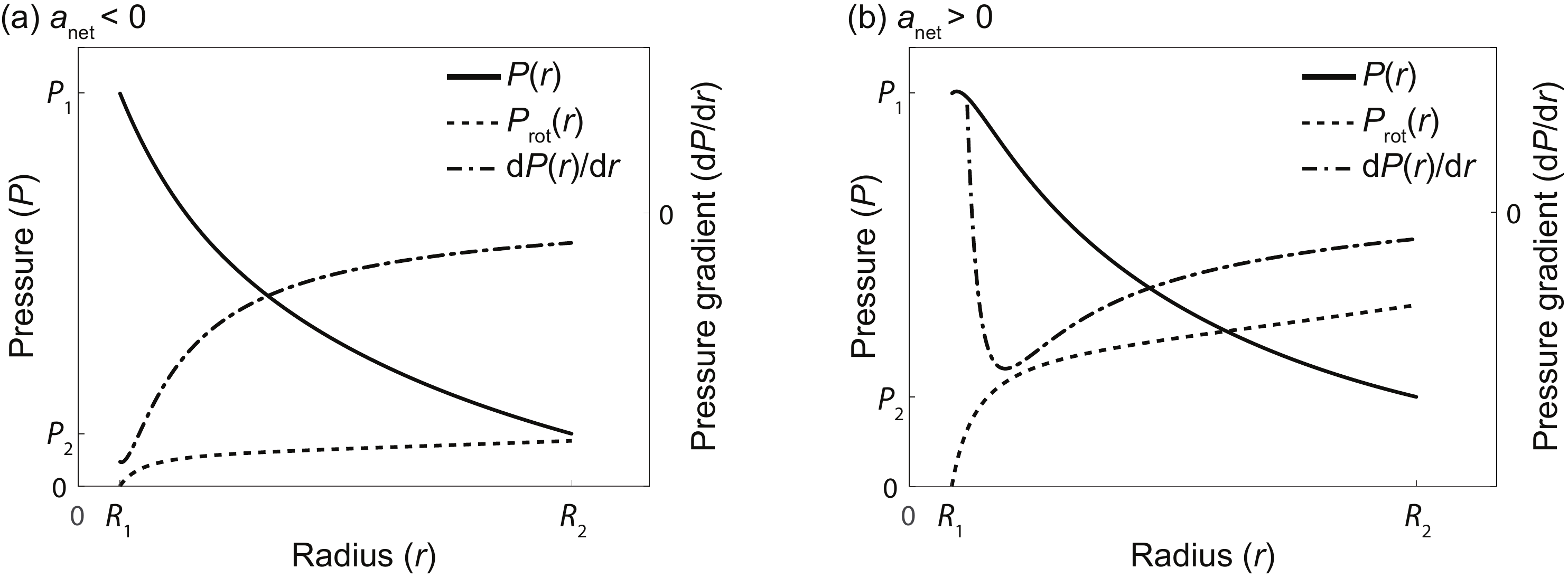}
\caption{Comparison of the pressure profile and pressure gradient within a cylindrically imploding liner at turnaround for (a) a non-stabilised rotating shell where the centripetal acceleration at the surface is lower than the radial acceleration, and (b) a stabilised rotating shell where the centripetal acceleration at the surface is greater than the radial acceleration. In (b) the rotation is sufficient to reverse the pressure gradient at the inner surface ($R_1$).}
\label{fig:15}
\end{figure}

The model developed in section~\ref{sec:4.0} and summarised by equation~\ref{eq:4:20} demonstrates that fluid rotation can stabilise the RT instability at the inner surface of an imploding liquid shell. The RT instability is induced by opposed pressure and density gradients at a fluid interface, as is typically the case at turnaround in a shell implosion. The centripetal acceleration that results from rotation contributes a positive (outwardly increasing) pressure gradient to the fluid, thus reducing the driving force for the RT instability, which can be entirely stabilised if rotation is sufficient to reverse the pressure gradient at the cavity surface. This effect is illustrated schematically in figure~\ref{fig:15}, where the radial distribution of the pressure and pressure gradient within a cylindrical liner at turnaround is compared for a rotating shell where the surface has an inward-facing net acceleration (i.e., the centripetal acceleration is greater than the radial acceleration), to a case with a lower initial angular velocity where there is a net outward acceleration. The contribution of rotation ($P_\mathrm{rot}(r)$) to the fluid pressure is also shown to illustrate the stabilisation mechanism. The pressure profiles were obtained using equation~\ref{eq:a:19}. As can be seen, the large gradient in the rotational pressure near the inner surface ($R_1$) reverses the total pressure gradient at the cavity surface for the case where the net acceleration is inward facing, which should stabilise the RT instability. These findings are in agreement with the analytical model of \citet{Avital2019}, which indicated that a net inwards facing interface acceleration at turnaround is required to stabilise the RT instability.

The experiments described in section~\ref{sec:5.0} were aimed at experimentally demonstrating the effect of rotation on the RT instability at the inner surface of a cylindrically imploding shell. The parameterisation of the unperturbed motion of the interface presented in section~\ref{sec:3.1} was used to select initial angular velocities and driving conditions that would produce geometrically similar implosions with significantly different ratios of centripetal acceleration to radial acceleration at turnaround. The unperturbed experiments, presented in section~\ref{sec:5.1}, were able to observe whether the inner surface of a cylindrically imploding shell would remain RT stable if the net acceleration is inward facing at turnaround, as was proposed in section~\ref{sec:4.0}. The low-mode-number experiments presented in section~\ref{sec:5.2} allowed for the direct measurement of the time varying perturbation amplitudes during the implosion. These measurements could be directly compared to equation~\ref{eq:4:20}, which is a model for the time evolution of azimuthal perturbations on the surface of cylindrically imploding shell initially in solid body rotation.  

\subsection{Unperturbed experiments}
\label{sec:6.1}

The off-axis images of the unperturbed experiments, presented in figure ~\ref{fig:11}, show a clear correlation between the initial angular velocity of the liquid shell and the appearance of high wavenumber perturbations (i.e., spray) on the cavity surface. More precisely, the experiments showed that the surface of the cavity remained stable as long as the net acceleration was inward facing and the condition for stabilising the RT instability, according to equation~\ref{eq:4:20}, was satisfied. For the 79~rad/s experiment, high wavenumber perturbations appeared on the surface of the cavity shortly following the point where the direction of the net acceleration transitioned from inward to outward facing. Increasing the initial angular velocity to 105~rad/s caused a delay in the appearance of perturbations and reduced the magnitude of the visible spray, while perturbation growth was entirely suppressed in the 131~rad/s experiment. The reduction in the magnitude of the outward facing net acceleration with increasing initial angular velocity was therefore seen to stabilise the shell inner surface from RT-driven perturbation growth. This result is consistent with images of stabilised implosions published by \citet{Turchi1980}. In general, the results of the unperturbed experiments indicate that any implosion where the net acceleration of the cavity surface remains inward facing during the implosion, or equivalently, where the ratio of the centripetal acceleration to radial acceleration at turnaround is greater than unity ($-a_{\mathrm{c}_\mathrm{max}}/ R_{1_\mathrm{max}}>1$, see parameter map of figure~\ref{fig:6}), should be stabilised from RT-driven perturbation growth.

The size of the perturbations which formed on the nominally smooth surface of the unperturbed experiments is related to the surface tension ($\sigma$) of the liquid as well as the wavenumber dependence of the RT instability. As can be seen in equation~\ref{eq:4:21}, in the absence of surface tension, the RT growth rate is proportional to the square root of the wavenumber ($k^{1/2}$) of the perturbations. The stabilising effect of surface tension, which is proportional to $k^{3/2}$, reduces the growth rate of short wavelength perturbations and completely stabilises sufficiently small features \citep{Chandrasekhar1961}. For a planar interface, the wavenumber which corresponds to the perturbation scale with the greatest growth rate can be expressed as \citep{Chandrasekhar1961}

\begin{equation}
\label{eq:6:1}
k_\mathrm{max}=\left[\frac{(\rho_2-\rho_1)g}{3\sigma}\right]^{\frac{1}{2}},
\end{equation}

\noindent where $g$ is the acceleration of the interface in the direction of the dense fluid ($\rho_2$). Equation~\ref{eq:6:1} should provide a reasonable estimate of $k_\mathrm{max}$ for the experimental conditions considered in this work, provided that $k_\mathrm{max}R_1 \gg 1$. Taking $g = 10$~mm/ms$^2$ as an order of estimate of the average net acceleration when the surface goes unstable, the predicted wavelength which maximizes perturbation growth is 0.9~mm, which correlates well with the scale of the perturbations observed in the 79~rad/s experiment(\SIrange{0.5}{2}{\milli\metre}). It is interesting to note that for marginally unstable implosions, where the magnitude and duration of the outward-facing net acceleration at turnaround are sufficiently small, surface tension may effectively stabilise the interface by limiting the perturbation growth rate. More precisely, the liquid shell inner surface will remain nominally smooth if the product of the perturbation growth rate for $k_\mathrm{max}$ and the duration of the interface net outward acceleration is on the order of or less than unity: $\gamma_\mathrm{max}\dot{R}_{1_\mathrm{max}}/\ddot{R}_{1_\mathrm{max}} \leq 1$, where the $\dot{R}_{1_\mathrm{max}}/\ddot{R}_{1_\mathrm{max}}$ factor is used to estimate the timescale of the net outwards acceleration. If this condition is met, the RT growth rate is insufficient to cause significant perturbation growth at the liquid shell inner surface during the period of interface instability near turnaround, thus effectively stabilising the interface. This mechanism may explain the notable difference in the amount of surface spray observed in the 79~rad/s and 105~rad/s experiment, despite both implosions being RT unstable.

\subsection{Low-mode-number experiments}
\label{sec:6.2}

In the low-mode-number experiments, an obstacle plate allowed for the formation of perturbations which had large sizes and modest growth rates, such that the evolution in their amplitude could be tracked and compared to the model predictions of equation~\ref{eq:4:20}. While the low-mode-number experiments used the same initial rotation rates and driving conditions as the unperturbed experiments, the presence of the obstruction plate fins partially obstructed the rotation of the flow, which directed the flow inward and resulted in additional convergence. This effect became more notable as the initial angular velocity was increased and can clearly be seen for the 131~rad/s experiment in figure~\ref{fig:12}(c), where the final radius at turnaround was 1.2~mm smaller than that of the model verification experiment. Despite this effect, the acceleration-time plots of figure~\ref{fig:12}(d)-(f) show that the desired variation in the net acceleration profile of the cavity surface was achieved for the three initial rotation rates, again producing a decrease in the net outward acceleration at turnaround with increasing initial angular velocity, and a net inward acceleration through turnaround for the 131~rad/s experiment.

The evolution in the perturbation amplitude observed in the mode-6 experiments showed good agreement with the model of equation~\ref{eq:4:20}, indicating that the relatively simple model captures a number of the important effects observed in the implosion. As is evident from equation~\ref{eq:4:20}, the evolution of perturbations on a cylindrically imploding surface is coupled to the bulk motion of the interface ($R_1$, $\dot{ R}_1$, $\ddot{R}_1$). A positive (inward facing) net surface acceleration results in the stable oscillation of perturbations on the interface (gravity waves): the first and last term of equation~\ref{eq:4:20} form a simple harmonic motion differential equation. This oscillating behaviour explains the phase reversal observed in all experiments at a radius of approximately 35~mm, and has been observed previously in experiments \citep{Weir1998} and models \citep{Barcilon1974,Mikaelian2005,Avital2019}. The oscillation period depends on the acceleration, radius, and mode number of the perturbations. Movie~6 of the supplementary material shows an implosion of a cavity with a mode-24 perturbation with the same initial angular velocity and driving conditions as the 131~rad/s experiments, where multiple phase reversals were observed during the implosion. Despite the apparent stability of these surface oscillations, convergence effects (i.e., Bell-Plesset effects, see $1/R_1$ factors in equation~\ref{eq:4:20}) cause the average perturbation amplitude to grow as the implosion proceeds \citep{Bell1951,Plesset1954}. This effect was clearly observed in the 131~rad/s experiment, where the amplitude of the perturbations increased from 0.8~mm at a radius of 55~mm, when the first phase reversal began, to 1.6~mm at a radius of 15~mm where the perturbations began a second phase reversal. It is interesting to note that, as can be seen in figure~\ref{fig:14}, the perturbation growth was very similar between experiments for the majority of the implosion (\SIrange{0}{4}{\milli\second}, \SIrange{74}{21}{\milli\metre}), despite the large variation in the initial angular velocity. During the majority of the implosion, the magnitude of the inward-facing radial acceleration at the cavity surface was much greater than the centripetal acceleration, and as a result the net acceleration at the cavity surface was similar for all three experiments (see figure~\ref{fig:8}(c)). It was only near the end of the implosion, where the conservation of angular momentum had amplified the difference in the cavity surface centripetal acceleration between experiments, that a notable difference in the perturbation growth was observed. 

Near turnaround, the initial angular velocity of the fluid had a significant effect on the net acceleration at the cavity surface, which resulted in a reduction in perturbation amplitude with increasing $\omega_0$ (see figure~\ref{fig:13}). The sudden difference in the evolution of the perturbation amplitude between the experiments which occurred beyond a radius of approximately 21~mm (see figure~\ref{fig:14}), coincided with a reversal in the direction of the net acceleration for the 79~rad/s and 105~rad/s tests, and can therefore be attributed to RT instability growth at the two lower angular velocities. The RT-instability-driven growth of low-mode-number perturbations on the inner surface of a cylindrically imploding shell near turnaround has been observed experimentally \citep{Weir1998} for non-rotating shells and theoretically for both rotating \citep{Barcilon1974} and non-rotating shells \citep{Mikaelian2005}. The observed difference in the amplitude of the perturbations near turnaround between the 79~rad/s and 105~rad/s experiments (see figures~\ref{fig:13} and~\ref{fig:14}) can be attributed to a reduction in the magnitude of the outward facing net acceleration with increasing $\omega_0$, which resulted in lower perturbation growth rates for the 105~rad/s test (see equation~\ref{eq:4:21}). For the 131~rad/s experiment, the growth in the perturbation amplitude was seen to level off and slightly decrease just prior to turnaround as it began to undergo a second phase reversal, demonstrating complete stabilisation of the RT instability. In general, it should be expected that for implosions where the net acceleration at the cavity surface becomes outward facing near turnaround, low-mode-number perturbations will undergo significant growth due to the RT instability during the final stages of implosion. For implosions where the net acceleration remains inward facing through turnaround, the surface is RT stable and the amplitude of perturbations should continue to evolve based on the surface oscillation and convergence effects discussed above for the early implosion.

Although there is good agreement between the low-mode-number experiments and the model of equation~\ref{eq:4:20}, there are important effects which occur near turnaround that the simplified one-dimensional model cannot capture. First, the large amplitude of the perturbations means that the linearisation assumptions made in deriving equation~\ref{eq:4:20} no longer apply. The perturbation amplitude which corresponds to the limit of the linear regime, defined as 10\% of the wavelength, is 1.6~mm for a minimum radius of 15~mm, a threshold that was surpassed by the 79~rad/s and 105~rad/s experiments. It is well known that the amplitude of perturbations driven by the RT instability transitions from exponential to linear growth as perturbations surpass the amplitude where the linear equations are applicable \citep{Haan1989,Goncharov2002}, which likely contributed to the observation that the model predicted perturbation growth rate in figure~\ref{fig:14} exceeded the experimentally observed growth rate for the 79~rad/s and 105~rad/s experiments near turnaround. Another effect that is not considered by the model of section~\ref{sec:4.0} is the distortion of the perturbation shape due to the Coriolis acceleration of the fluid which induces a large angular velocity gradient ($\partial \omega$/$\partial r$) at the cavity surface near turnaround. The effect of the velocity shear on the interface can be seen in the images of figure~\ref{fig:13}, where the perturbations, which have a quasi-sinusoidal shape at a radius of 21~mm, become gradually more distorted as the implosion proceeds. This effect does not appear to be caused by a velocity shear between the liquid shell and the gas filled cavity (i.e., the Kelvin-Helmholtz instability), but rather by a gradient in the angular velocity of the liquid shell induced by the Coriolis acceleration. The angular velocity gradient within the liquid near the cavity surface caused the fluid at the spike of the perturbation to rotate faster than the fluid at the trough of the bubble, leading the perturbations to evolve into that of a breaking wave shape. As was discussed in section~\ref{sec:4.0}, Coriolis acceleration effects are not treated by equation~\ref{eq:4:20}, which assumes that the perturbations maintain sinusoidal symmetry. Based on the mechanism discussed above, the magnitude of perturbation distortion should depend on their amplitude and the angular velocity gradient near the cavity surface, which explains why it is only observed near turnaround. Modelling the complete motion of the perturbations, including Coriolis acceleration effects, for a time varying interface radius, would require a significantly more complicated theoretical model than that of section~\ref{sec:4.0}, but could readily be accomplished using hydrocode simulations. 

The distortion of the perturbations due to the velocity gradient meant that their amplitude could not be meaningfully tracked beyond turnaround and eventually caused the tip of the spikes to collide with an adjacent portion of the cavity inner surface. The phenomenon also lead to a narrowing of the tips of the perturbation spikes, which were observed to break-up into a spray after turnaround for the 79~rad/s and 105~rad/s experiments (see supplementary movie~4 and movie~5). Perhaps the most notable consequence of the wave breaking phenomenon is the disruption the quasi-reversible implosion behaviour predicted by the model of equation~\ref{eq:4:20} for RT stable implosions. Without this effect, low-mode-number perturbations would be expected to continue to undergo stable surface oscillations through turnaround and during the expansion phase, as long as the net acceleration remains inward facing. As can be seen in supplementary movies~4, 5, and~6, the distortion of the low-mode-number perturbations in the 131~rad/s experiments disrupted the interface at turnaround, so that the reversible expansion was not achieved. The experiments thus demonstrated that even for RT stable implosions, low-mode-number perturbations can reach a sufficient amplitude such that wave-breaking effects obstruct the expected stable phase reversal and cause spray to be ejected into the cavity, which could have devastating effects on plasma compression in an MTF reactor.

Other effects, not considered by the model of section~\ref{sec:4.0}, may have contributed to the observed stabilisation of mode-6 perturbations with increasing initial angular velocity. Theoretical models \citep{Mjolsness1986,Shumlak1995,Shumlak1998,Ruden2002} and Z-pinch plasma experiments \citep{Shumlak2001} have shown that velocity shear normal to the density gradient at a fluid interface can stabilise RT-driven perturbation growth. Similarly, in centrifugally confined magnetic-mirror plasma devices, velocity shear in the azimuthal plasma flow has been shown to stabilise the flute instability \citep{Huang2001,Teodorescu2005,Reid2014}. In the experiments presented in this work, conservation of angular momentum generates significant velocity shear in the radial direction ($\partial\omega /\partial r$) at the cavity surface, particularly near turnaround where the cavity surface may be subject to the RT instability. Shear stabilisation is most effective for wavelengths on the order of the shear layer thickness \citep{Mjolsness1986,Ruden2002}, due to the fact that the magnitude of shear required to stabilise the interface increases with increasing wavenumber, but shear stabilisation does not stabilise wavelengths that are larger than twice the thickness of the shear layer. The key parameter that determines the level of stabilisation is the ratio of the interface acceleration to flow shear, quantified by the Richardson number: $Ri=gd/U^2$, where $g$ is the interface acceleration (net acceleration in experiments), $d$ the thickness of the shear layer, and $U$ is the velocity difference across the shear layer. Models for shear stabilisation in a planar geometry with a linear velocity gradient in the shear layer have shown that $Ri$ must typically be smaller than 0.1 to stabilise the RT instability at wavelengths on the order of the shear layer thickness, and much smaller than 0.1 to stabilise a wide range of wavelengths \citep{Mjolsness1986,Ruden2002}. These results can be used to estimate the effect of shear on the mode-6 experiments. For the 105~rad/s experiment at $R_1$~=~20~mm: $g\approx$~\SI{e4}{\metre\per\second\squared}, $d\approx$~20~mm (estimated to be on the order of the wavelength), $U$~=~\SI{e3}{\metre\per\second}, which gives a Richardson number of $Ri$~=~0.2. It therefore does not appear likely that shear provided meaningful stabilisation in the mode-6 experiments. It would be possible to choose experimental settings that would ensure marginal stability ($- a_\mathrm{net}<$~\SI{e3}{\metre\per\second\squared}), where the shear would be sufficient to stabilise the RT instability, but even in this case, the contribution of the centripetal acceleration to stabilisation, which reduces $- a_\mathrm{net}$ from \SI{e5}{\metre\per\second\squared} for a non-rotating experiment to \SI{e3}{\metre\per\second\squared} is much more significant than the stabilising effect of shear.

Another effect which was not considered in the above discussion is the stabilisation of the interface due to the Coriolis force in the radial direction \citep{Tao2013,Scase2018}. As was discussed in section~\ref{sec:4.0}, velocity perturbations induce a flow in the azimuthal direction, which, in a rotating reference frame, is subject to the Coriolis force. It has been shown, using a model developed for the time evolution of azimuthal perturbations at a cylindrical fluid interface in solid body rotation at a fixed radius, that, in the high Atwood number limit, the linear growth rate of perturbations due to the RT instability should be reduced by the square of the angular velocity of the fluid at the interface \citep{Tao2013,Scase2018}. The linear RT growth rate for perturbations on the shell inner surface, including the effect of Coriolis stabilisation, can be expressed as \citep{Tao2013}

\begin{equation}
\label{eq:6:2}
\gamma=\left(\frac{M\ddot{R}_1}{R_1}-\frac{ M\omega_0^2R_{1_0}^4}{R_1^4}-\frac{ \omega_0^2R_{1_0}^4}{R_1^4}\right)^{\frac{1}{2}},
\end{equation}

\noindent where the first term is due to the radial acceleration of the cavity (positive near turnaround), the second term is the stabilising effect of the centripetal acceleration, and the final term is the stabilising effect of the Coriolis acceleration. As can be seen, for the flow considered in this work, the mode number ($M$) factor in the centripetal acceleration term causes the Coriolis acceleration effect to be small compared to the effect of the centripetal acceleration for all but very small mode number perturbations. 

\section{Conclusion}
\label{sec:7.0}
The stabilising effect of centripetal acceleration on the RT instability at the inner surface of a cylindrically imploding liquid shell initially in solid body rotation was studied experimentally and theoretically. A simple one-dimensional model for the evolution of perturbations on the inner surface of a rotating cylindrically imploding shell was developed, which demonstrated that the RT instability could be stabilised by the centripetal acceleration of the fluid, as long as the net acceleration of the cavity surface remained inward facing through turnaround. A series of experiments were performed using an apparatus that was capable of generating rotating shell implosions in a cylindrical geometry and allowed for visualisation of the implosion process. Test conditions were selected using a one-dimensional model for the unperturbed motion of the cavity surface, such that the implosions had similar convergence ratios and radial accelerations at turnaround, but significantly different net inner surface accelerations due to a variation in the initial angular velocity of the liquid shell. The experiments demonstrated that the rotation of the fluid on the inner surface of the shell could inhibit the RT-driven growth of high-mode-number and low-mode-number perturbations which would otherwise be caused by the large outward radial acceleration of the surface near turnaround. As was predicted by the analytical model, the RT-instability-driven growth of perturbations that was observed in experiments at low initial angular velocities was supressed when the rate of rotation was increased such that the net acceleration at the shell inner surface was inwards facing at turnaround. Good agreement was seen between the experimentally observed evolution in the amplitude of the mode-6 perturbations and the simple analytical model, indicating that for the conditions probed in this work, the centripetal acceleration of the fluid is the dominant mechanism by which rotation affects perturbation growth. The low-mode-number experiments showed that the significant angular velocity gradient near the shell inner surface at turnaround can distort the perturbations and cause the impact of the spikes with an adjacent portion of the cavity surface, an effect which may be of significant concern for plasma compression applications.

\section*{Acknowledgements} 
The authors would like to thank Victoria Suponitsky for her guidance with the experiments and analysis, as well as useful feedback on the manuscript. The contributions of the following individuals in developing the apparatus and performing experiments is also greatly appreciated: Takiah Ebbs-Picken, Zhe Ding, Hansen Liu, Zhuo Fan Bao, Jihane Kamil, Hin Fung Ng, Samuël Vaillancourt, Adamo Albanese, Gordon Faust, Robert Laniel, Jennifer Kwiatkowski, Yann Beaulieu, Philippe Toren, Douglas Prisnie, David King-Hope, Charles Calnan, and Zain Eejaz. This work was supported by General Fusion and the Natural Sciences and Engineering Research Council of Canada (NSERC) under Collaborative Research and Development Grant 477617-14.

\appendix
\section{}
\label{app:A}

\begin{figure}
	\centering
	\includegraphics[width=0.3\columnwidth]{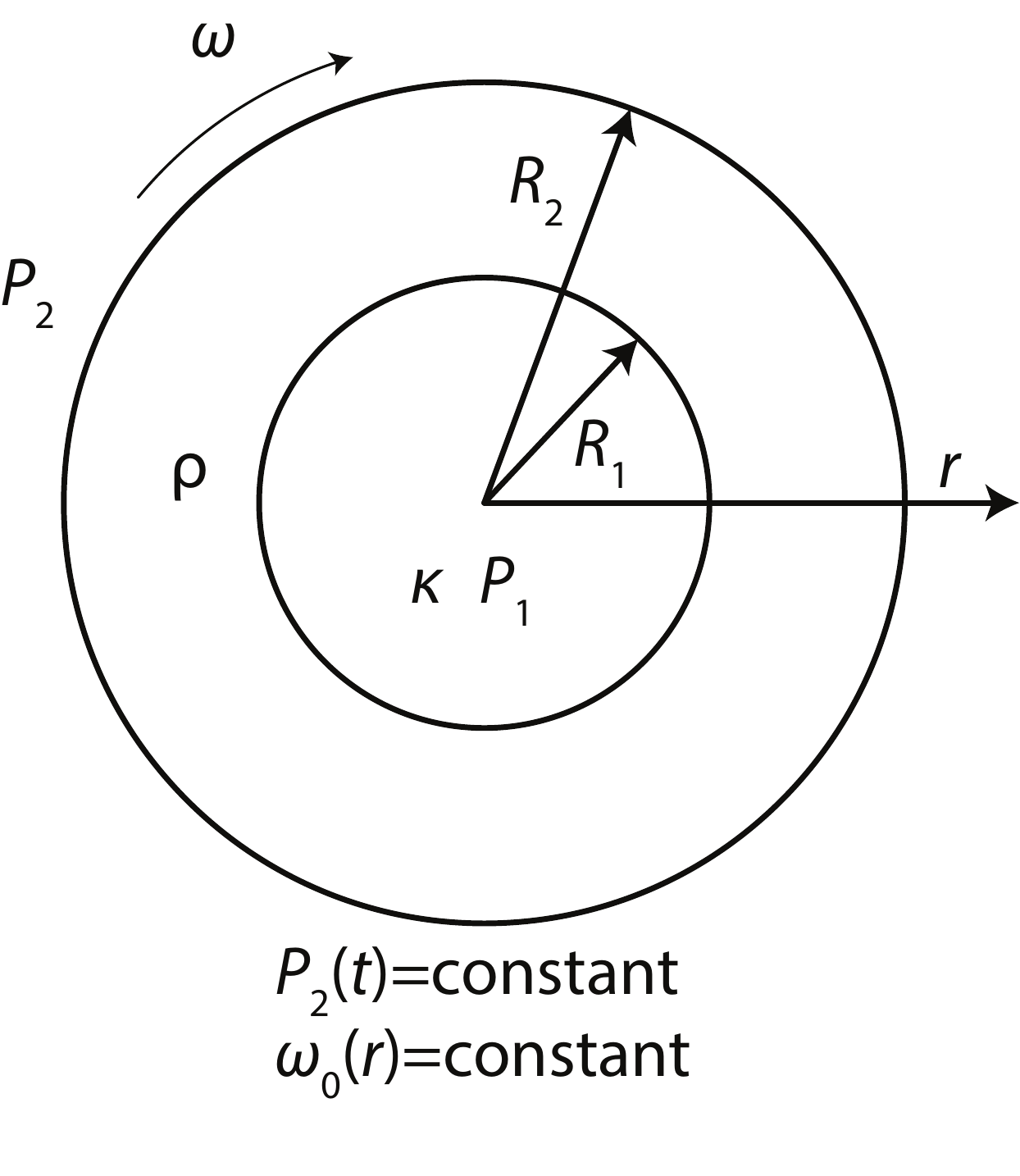}
	\caption{Schematic of the problem definition for the development of a model for the one-dimensional motion of a rotating cylindrically imploding shell.}
	\label{fig:16}
\end{figure}

The energy conservation approach described in section~\ref{sec:3.05} will be used to derive an analytic expression relating the radius and radial velocity of the inner surface of a rotating cylindrical shell. The system, pictured in figure~\ref{fig:16}, consists of a fluid having an inner radius ($R_1$) and an outer radius ($R_2$), initially in solid body rotation ($\omega_0$). The cavity is filled with a calorically perfect gas with a specific heat ratio ($\kappa$) which undergoes isentropic compression. The implosion will be driven by a constant pressure ($P_2$) acting on the outer surface of the shell. The shell fluid will be assumed incompressible (constant density, $\rho$) and inviscid and its angular momentum will be conserved during the implosion. The subscript ``$_0$'' will be used to denote initial conditions. The conservation of energy for the system can be expressed by the following equation

\begin{equation}
\label{eq:a:1}
W_ {\mathrm{in}}=\Delta K_ {\mathrm{radial}}+\Delta K_ {\mathrm{rotational}}+\Delta I_ {\mathrm{gas}},
\end{equation}

\noindent where the work done on the system by the external pressure ($W_ {\mathrm{in}}$) is equal to the sum of the change in the radial kinetic energy of the shell ($K_ {\mathrm{radial}}$), the rotational kinetic energy of the shell ($K_ {\mathrm{rotational}}$), and the internal energy of the gas within the cavity ($I_ {\mathrm{gas}}$). Equation~\ref{eq:a:1} will be used to develop an expression for the radial velocity of the shell inner surface ($\dot{R}_1$) as a function of its radius which will define the motion of the shell. Equation~\ref{eq:a:1} must therefore be expressed in terms of the position and radial velocity of the interface and initial conditions alone. The terms will be normalised by the height of the cylindrical system which has no effect on the final expression. The work done on the system by the external pressure can be determined by the following integral:

\begin{equation}
\label{eq:a:2}
W_ {\mathrm{in}}=\int_{V_0}^{V}P_2(t)\mathrm{d}V,
\end{equation}

\noindent where $\mathrm{d}V$ is the change in the external volume and, equivalently, the volume of the inner cavity for an incompressible system. For the case of a constant external pressure, the work done on the system is

\begin{equation}
\label{eq:a:3}
W_ {\mathrm{in}}=\pi P_2(R_{1_0}^2-R_1^2).
\end{equation}

\noindent The change in the internal energy of the gas is defined by the following expression

\begin{equation}
\label{eq:a:4}
\Delta I_ {\mathrm{gas}}=\frac{P_1V_1-P_{1_0}V_{1_0}}{\kappa-1},
\end{equation}

\noindent where $P_1$ and $V_1$ are the pressure of the cavity gas and the volume of the cavity, respectively. For a gas undergoing isentropic compression, the pressure can be related to the inner radius of the shell:

\begin{equation}
\label{eq:a:5}
P_1=P_{1_0}\Big(\frac{R_{1_0}}{R_1}\Big)^{2\kappa}.
\end{equation}

\noindent Combining equations~\ref{eq:a:4} and~\ref{eq:a:5}, the change in the gas internal energy can be expressed as follows:

\begin{equation}
\label{eq:a:6}
\Delta I_ {\mathrm{gas}}=\frac{\pi P_{1_0}R_{1_0}^2}{\kappa-1}\bigg[\Big(\frac{R_{1_0}}{R_1}\Big)^{2(\kappa -1)}-1\bigg].
\end{equation}

\noindent The change in the radial kinetic energy of the shell, which has no initial radial velocity, can be evaluated by the following integral

\begin{equation}
\label{eq:a:7}
\Delta K_ {\mathrm{radial}}=\int_{R_1}^{R_2}\frac{1}{2}\dot{r}^2 \mathrm{d}m,
\end{equation}

\noindent where $\dot{r}$ is the radial velocity of the differential element $\mathrm{d}m=2\pi \rho r \mathrm{d}r$. For an incompressible fluid, the volume flux in the radial direction must be equal to the volume flux at the cavity surface so that the radial velocity can be related to the shell inner surface radius and velocity in the following manner:

\begin{equation}
\label{eq:a:8}
\dot{r}=\frac{\dot{R}_1 R_1}{r}.
\end{equation}

\noindent The integral of equation~\ref{eq:a:7} can be re-written as

\begin{equation}
\label{eq:a:9}
\Delta K_ {\mathrm{radial}}=\pi \rho \dot{R}_1^2 R_{1}^{2}\int_{R_1}^{R_2}\frac{1}{r}\mathrm{d}r.
\end{equation}

\noindent From conservation of volume, the outer radius of the shell can be obtained from the change in the radius of the inner surface:

\begin{equation}
\label{eq:a:10}
R_2=(R_{2_0}^2-R_{1_0}^2+R_1^2)^{\frac{1}{2}}.
\end{equation}

\noindent Finally, the integral of equation~\ref{eq:a:9} can be evaluated:

\begin{equation}
\label{eq:a:11}
\Delta K_ {\mathrm{radial}}=\pi \rho \dot{R}_1^2 R_{1}^{2}\,\mathrm{ln}\left[\frac{(R_{2_0}^2-R_{1_0}^2+R_1^2)^{\frac{1}{2}}}{R_1}\right].
\end{equation}

\noindent The rotational kinetic energy of the shell can be obtained from the following integral:

\begin{equation}
\label{eq:a:12}
K_ {\mathrm{rotational}}=\int_{R_1}^{R_2}\frac{1}{2}r^2\omega^2\mathrm{d}m,
\end{equation}

\noindent where $r$ and $\omega$ are the radial position and angular velocity of the differential element $\mathrm{d}m=2\pi \rho r \mathrm{d}r$. The initial rotational kinetic energy can be obtained by directly evaluating the integral of equation~\ref{eq:a:12}:

\begin{equation}
\label{eq:a:13}
K_ {\mathrm{rotational}}\big|_{t=0}=\frac{\pi}{4}\rho\omega_0^2(R_{2_0}^4-R_{1_0}^4).
\end{equation}

\noindent As the implosion proceeds, the conservation of angular momentum creates an angular velocity gradient within the shell. The angular velocity at position $r$ can be related to the initial angular velocity and position of the fluid element:

\begin{equation}
\label{eq:a:14}
\omega=\omega_0\Big(\frac{r_0}{r}\Big)^2.
\end{equation}

\noindent From the conservation of mass, the initial position of the fluid element can be related to its current position and the change in the position of the cavity surface:

\begin{equation}
\label{eq:a:15}
r_0=(R_{1_0}^2-R_{1}^2+r^2)^{\frac{1}{2}}.
\end{equation}

\noindent The integral of equation~\ref{eq:a:12} can be re-written as:

\begin{equation}
\label{eq:a:16}
K_ {\mathrm{rotational}}=\pi\rho\omega_0^2\int_{R_1}^{R_2}r^3+2r(R_{1_0}^2-R_1^2)+\frac{1}{r}(R_{1_0}^2-R_1^2)^2\:\mathrm{d}r.
\end{equation}

\noindent The following expression for the rotational kinetic energy of the liner can be obtained by evaluating the integral and substituting equation~\ref{eq:a:10} for the outer radius:

\begin{eqnarray}
\label{eq:a:17}
K_ {\mathrm{rotational}}=\pi\rho\omega_0^2\bigg[\frac{(R_{2_0}^2-R_{1_0}^2)^2+2R_1^2(R_{2_0}^2-R_{1_0}^2)}{4}+(R_{2_0}^2-R_{1_0}^2)(R_{1_0}^2-R_1^2)\nonumber \\
+\mathrm{ln}\bigg(\frac{(R_{2_0}^2-R_{1_0}^2+R_1^2)^{\frac{1}{2}}}{R_1}\bigg)(R_{1_0}^2-R_{1}^2)^2\bigg].
\end{eqnarray}

\noindent The change in the rotational kinetic energy is obtained from the difference between equations~\ref{eq:a:17} and~\ref{eq:a:13}:

\begin{eqnarray}
\label{eq:a:18}
\Delta K_ {\mathrm{rotational}}=\pi\rho\omega_0^2\bigg[\frac{R_{1_0}^4-R_{2_0}^2R_{1_0}^2+R_1^2(R_{2_0}^2-R_{1_0}^2)}{2}+(R_{2_0}^2-R_{1_0}^2)(R_{1_0}^2-R_1^2)\nonumber \\
+\mathrm{ln}\bigg(\frac{(R_{2_0}^2-R_{1_0}^2+R_1^2)^{\frac{1}{2}}}{R_1}\bigg)(R_{1_0}^2-R_{1}^2)^2\bigg].
\end{eqnarray}

\noindent Equations~\ref{eq:a:3}, \ref{eq:a:6}, \ref{eq:a:11}, and~\ref{eq:a:18} can be substituted into equation~\ref{eq:a:1} to find an expression for the radial velocity of the shell inner surface as a function of its radius. As explained in section~\ref{sec:3.05}, the time evolution of the shell radius, velocity, and acceleration can then be obtained using finite difference methods. The approach presented above is equivalent to developing a differential equation for the time evolution of the shell inner surface based on unsteady potential flow \citep{Kull1991}. The motion of the shell inner surface for the system considered here is defined by the second-order ordinary differential equation below:

\begin{equation}
\label{eq:a:19}
(R_1\ddot{R}_1+\dot{R}_1^2)\mathrm{ln}\Big(\frac{R_2}{R_1}\Big)-\frac{1}{2}\dot{R}_1^2\Big(1-\frac{R_1^2}{R_2^2}\Big)=\frac{P_1-P_2+P_{\mathrm{rot}}}{\rho},
\end{equation}

\noindent where $R_2$ can again be related to $R_1$ and the initial conditions using equation~\ref{eq:a:10}. The $P_{\mathrm{rot}}$ term accounts for the pressure gradient induced by the rotation of the shell and can be obtained using the following integral:

\begin{equation}
\label{eq:a:20}
\frac{P_{\mathrm{rot}}}{\rho}=\int_{R_1}^{R_2}r\omega^2\mathrm{d}r.
\end{equation}

\noindent The expression for the pressure induced by rotation can be obtained by substituting equations~\ref{eq:a:14} and~\ref{eq:a:15} into equation~\ref{eq:a:20} and performing the integration:

\begin{equation}
\label{eq:a:23}
\frac{P_{\mathrm{rot}}}{\rho}=\omega_0^2\Big[\frac{R_2^2-R_1^2}{2}+2\mathrm{ln}\Big(\frac{R_2}{R_1}\Big)(R_{1_0}^2-R_1^2)-\frac{1}{2}\Big(\frac{1}{R_2^2}-\frac{1}{R_1^2}\Big)(R_1^4+R_{1_0}^4-2R_1^2R_{1_0}^2)\Big].
\end{equation}

\noindent Equations~\ref{eq:a:1} and~\ref{eq:a:19} are equivalent, however the energy approach presented above is readily adaptable to the shell geometry used in the experimental work presented in this paper.

\bibliography{RTbibliography}
\bibliographystyle{jfm}

\end{document}